\documentstyle[epsfig]{mn}

\newif\ifAMStwofonts

\DeclareMathAlphabet{\mathsc}{OT1}{cmr}{m}{sc}
\def\testbx{bx}%
\DeclareRobustCommand{\ion}[2]{%
\relax\ifmmode
\ifx\testbx\f@series
{\mathbf{#1\,\mathsc{#2}}}\else
{\mathrm{#1\,\mathsc{#2}}}\fi
\else\textup{#1\,{\mdseries\textsc{#2}}}%
\fi}

\ifoldfss
  \ifCUPmtlplainloaded \else
    \NewTextAlphabet{textbfit} {cmbxti10} {}
    \NewTextAlphabet{textbfss} {cmssbx10} {}
    \NewMathAlphabet{mathbfit} {cmbxti10} {} 
    \NewMathAlphabet{mathbfss} {cmssbx10} {} 
  \fi
  \ifAMStwofonts
    \ifCUPmtlplainloaded \else
      \NewSymbolFont{upmath} {eurm10}
      \NewSymbolFont{AMSa} {msam10}
      \NewMathSymbol{\upi}     {0}{upmath}{19}
      \NewMathSymbol{\umu}     {0}{upmath}{16}
      \NewMathSymbol{\upartial}{0}{upmath}{40}
      \NewMathSymbol{\leqslant}{3}{AMSa}{36}
      \NewMathSymbol{\geqslant}{3}{AMSa}{3E}

       \let\le=\leqslant
       \let\ge=\geqslant
    \fi
  \fi
\fi 

\ifnfssone
  \newmathalphabet{\mathit}
  \addtoversion{normal}{\mathit}{cmr}{m}{it}
  \addtoversion{bold}{\mathit}{cmr}{bx}{it}
  \newmathalphabet{\mathbfit} 
  \addtoversion{normal}{\mathbfit}{cmr}{bx}{it}
  \addtoversion{bold}{\mathbfit}{cmr}{bx}{it}
  \newmathalphabet{\mathbfss} 
  \addtoversion{normal}{\mathbfss}{cmss}{bx}{n}
  \addtoversion{bold}{\mathbfss}{cmss}{bx}{n}
  \ifAMStwofonts
    \ifCUPmtlplainloaded \else
      \UseAMStwoboldmath
      \makeatletter
      \new@mathgroup\upmath@group
      \define@mathgroup\mv@normal\upmath@group{eur}{m}{n}
      \define@mathgroup\mv@bold\upmath@group{eur}{b}{n}
      \edef\UPM{\hexnumber\upmath@group}
      \new@mathgroup\amsa@group
      \define@mathgroup\mv@normal\amsa@group{msa}{m}{n}
      \define@mathgroup\mv@bold\amsa@group{msa}{m}{n}
      \edef\AMSa{\hexnumber\amsa@group}
      \makeatother
      \mathchardef\upi="0\UPM19
      \mathchardef\umu="0\UPM16
      \mathchardef\upartial="0\UPM40
      \mathchardef\leqslant="3\AMSa36
      \mathchardef\geqslant="3\AMSa3E

       \let\le=\leqslant
       \let\ge=\geqslant
    \fi
  \fi
\fi 

\ifnfsstwo
  \DeclareMathAlphabet{\mathbfit}{OT1}{cmr}{bx}{it}
  \SetMathAlphabet\mathbfit{bold}{OT1}{cmr}{bx}{it}
  \DeclareMathAlphabet{\mathbfss}{OT1}{cmss}{bx}{n}
  \SetMathAlphabet\mathbfss{bold}{OT1}{cmss}{bx}{n}
  \ifAMStwofonts
    \ifCUPmtlplainloaded \else
      \DeclareSymbolFont{UPM}{U}{eur}{m}{n}
      \SetSymbolFont{UPM}{bold}{U}{eur}{b}{n}
      \DeclareSymbolFont{AMSa}{U}{msa}{m}{n}
      \DeclareMathSymbol{\upi}{0}{UPM}{"19}
      \DeclareMathSymbol{\umu}{0}{UPM}{"16}
      \DeclareMathSymbol{\upartial}{0}{UPM}{"40}
      \DeclareMathSymbol{\leqslant}{3}{AMSa}{"36}
      \DeclareMathSymbol{\geqslant}{3}{AMSa}{"3E}

       \let\le=\leqslant
       \let\ge=\geqslant
    \fi
  \fi
\fi 

\ifCUPmtlplainloaded \else
  \ifAMStwofonts \else 
    \def\upi{\pi}
    \def\umu{\mu}
    \def\upartial{\partial}
  \fi
\fi

\title[The Lyman Forest at $z > 1.5$]
{The Physical Properties of the Ly$\alpha$ Forest at $z > 1.5$\thanks{The
data used in this study are based on public
data released from the UVES Commissioning and Science
Verification and from the OPC program 65.O-0296A (P.I. S.
D'Odorico) at the VLT/Kueyen
telescope, ESO, Paranal, Chile.}$^{,}$\thanks{The line lists
in the study are only available in electronic form at the CDS 
via anonymous ftp
to cdsarc.u-strasbg.fr (130.79.128.5).}}
\author[T.-S. Kim et al.]
       {T.-S.~Kim,$^1$ R.~F.~Carswell,$^2$ S.~Cristiani,$^{3,4}$
       S.~D'Odorico$^1$ and E.~Giallongo$^{5}$ \\
       $^1$European Southern Observatory,
       Karl-Schwarzschild-Strasse 2, D-85748, Garching b.
       M\"unchen, Germany\\
       $^2$Institute of Astronomy, Madingley Road, Cambridge CB3 0HA,
       U.K.\\
       $^3$ST European Coordinating Facility, ESO,
       Karl-Schwarzschild-Strasse 2, D-85748, Garching b.
       M\"unchen, Germany\\
       $^4$Osservatorio Astronomico di Trieste, via G. B. Tiepolo 11,
       I-34131 Trieste, Italy\\
       $^5$Osservatorio Astronomico di Roma, via dell'Osservatorio 2,
       I-00040 Monteporzio, Italy}
\date{Accepted .
      Received ;
      in original form}

\pubyear{2002}

\begin{document}

\maketitle
\begin{abstract}
Combining a new, increased dataset of 8 QSOs covering the Ly$\alpha$
forest at redshifts $1.5 < z < 3.6$ from VLT/UVES observations with
previously published results, we have investigated the properties of
the Ly$\alpha$ forest at $1.5 < z < 4$. With the 6 QSOs covering
the Ly$\alpha$ forest at $1.5 < z < 2.5$, we have extended previous
studies in this redshift range. In particular, we have
concentrated on the evolution of the line number density and the
clustering of the Ly$\alpha$ forest at $z \le 2.5$, where the Ly$\alpha$
forest starts to show some inhomogeneity from sightline to sightline.
We have fitted Voigt profiles to the Ly$\alpha$ absorption lines as in
previous studies, and have, for two QSOs with $z_{\mathrm{em}} \sim
2.4$, fitted Ly$\alpha$ and higher order of Lyman lines down
to 3050 \AA\/ simultaneously. This latter approach 
has been taken in order to
study the Ly$\beta$ forest at $z \sim 2.2$ and the higher
\ion{H}{i} column density Ly$\alpha$ forest 
in the Ly$\beta$ forest region.

For a given $N_\ion{H}{i}$ range, the Ly$\alpha$ forest at $1.5 < z <
4$ shows the monotonic evolution, which is governed mainly by the
Hubble expansion at this redshift range. In general, the Ly$\alpha$ forest
line number density ($dn/dz$) is best approximated with $dn/dz = 6.1 \,
(1+z)^{2.47 \pm 0.18}$ for the \ion{H}{i} column density $N_\ion{H}{i}
= 10^{13.64 - 17} \ {\mathrm{cm}}^{-2}$ at $1.5 < z < 4$. 
When the results at $0 < z < 1.5$
from {\it HST} observations are combined, the slow-down in the number density
evolution occurs at $z < 1.5$. For higher column density clouds at 
$N_\ion{H}{i} > 10^{14}
\ {\mathrm{cm}}^{-2}$, there is a variation in
the line number density from sightline to sightline at $z < 2.5$.
This variation is stronger for higher column density systems, 
probably due to more gravitationally evolved structures 
at lower $z$. The mean \ion{H}{i} opacity
$\overline{\tau}_\ion{H}{i}$ is $\overline{\tau}_\ion{H}{i} (z) =
0.0032 \, (1+z)^{3.37 \pm 0.20}$ at $1.5 < z < 4$.   {\it HST} observations
show evidence for slower evolution of $\overline{\tau}_\ion{H}{i}$ at
$z < 1$.  For $N_\ion{H}{i} = 10^{12.5-15} \ {\mathrm{cm}^{-2}}$, the
differential column density distribution function, $f(N_\ion{H}{i})$,
can be best fit by $f(N_\ion{H}{i}) \propto N_{\ion{H}{i}}^{-\beta}$
with $\beta \approx 1.5$ for $1.5 < z < 4$.  When combined with {\it HST}
observations, the exponent $\beta$ increases as $z$ decreases at $0 < z
< 4$ for $N_\ion{H}{i} = 10^{13-17} \ {\mathrm{cm}^{-2}}$.  The
correlation strength of the step optical depth correlation function shows the
strong evolution from $<\!z\!> \ = 3.3$ to $<\!z\!> \ = 2.1$, although
there is a large scatter along different sightlines.  The analyses of
the Ly$\beta$ forest at $z \sim 2.2$ are, in general, in good agreement
with those of the Ly$\alpha$ forest.
\end{abstract}


\section{Introduction}
The redshift evolution of the Ly$\alpha$ forest imprinted in the
spectra of high-$z$ QSOs provides a powerful tool to probe the
distribution and evolution of baryonic matter, and hence the formation
and evolution of galaxies and the large scale structure, over a wide
range of redshifts up to $z \sim 6$ (Sargent et al. 1980; Schaye et
al.  1999; Kim, Cristiani \& D'Odorico 2001, hereafter KCD).

The evolution of the Ly$\alpha$ forest is mainly governed by two
physical processes. One is the Hubble expansion and another is the
ionizing ultraviolet background flux (Theuns, Leonard \& Efstathiou
1998; Dav\'e et al. 1999; Schaye et al. 2000; Bianchi, Cristiani \& Kim
2001). At higher $z$, the Hubble expansion and the non-decreasing
ultraviolet background cause a rapid evolution of the line number
density per unit redshift, $dn/dz$.  At lower $z$, the number of
photons available to ionize the Ly$\alpha$ forest becomes smaller, due
to the decrease of the number of QSOs at $z < 2$, which is generally
assumed to be the main source of the ionizing photons. As a result the
rate of change of $dn/dz$ with redshift is smaller.

Studies on the forest at $z > 2$ have shown the rapid evolution of the
line number density (Lu, Wolfe \& Turnshek 1991; Bechtold 1994; Kim et
al. 1997; KCD). Weymann et al. (1998) from the {\it HST} QSO
absorption line key project, however, have shown that the redshift
evolution of the line number density is much more gradual at redshifts
below $z \sim 1.5$ than above. These results suggest that the
transition between the two different evolutionary rates occur somewhere
in the range $z \sim 1.3 \to 1.7$. From numerical simulations, Theuns
et al. (1998) and Dav\'e et al. (1999) have demonstrated
that the change in the evolutionary slope occurs at $z \sim 1.7$ due to
the decreasing UV background at $z  < 2$, assuming a QSO-dominated
background.  Recently, from the study of the Ly$\alpha$ forest at $1.5
< z < 2.4$, KCD conclude that the Ly$\alpha$ forest at $1.5 < z < 4$
shows a monotonic, continuous evolution with $z$, in terms of the line
number density, the mean \ion{H}{i} opacity and the correlation
strength, both from the profile fitting analysis and from the optical
depth analysis. KCD also show that the change in the number density
evolution occurs at $z \sim 1.2$, suggesting that a QSO-dominated UV
background used in numerical simulations underestimates the UV
background and/or the enhanced structure formation at $z < 2.5$ (Theuns
et al. 1998; Weymann et al. 1998; Dav\'e et al. 1999; Bianchi et al.
2001).  Therefore, it is of importance to investigate the redshift
range where $dn/dz$ starts to change and to study any variations from
sightline to sightline from larger statistical sample in order to
constrain the results from numerical simulations and the nature of the
ionizing sources.

Here, we present the analysis of six QSOs covering the Ly$\alpha$
forest at $1.5 < z_{\mathrm{Ly\alpha}} < 2.4$ as well as two QSOs at $3
< z_{\mathrm{Ly\alpha}} < 3.6$ from  high resolution ($R \sim
45\,000$), high S/N ($\sim$ 35--50) spectra obtained with the
VLT/UVES.  This analysis is an extension of the one by
KCD with a larger uniform sample.
We consider the evolutionary behavior of the Lyman forest
systems over the redshift range $1.5 < z < 2.4$, in particular the line
number density evolution and the clustering, and compare the results
with those found by others at higher and lower redshifts. We have
adopted the traditional Voigt profile fitting method to analyse the
absorption lines in two ways. In the first approach, we have fitted the
Ly$\alpha$ forest region using the Ly$\alpha$ lines only, since most
previous studies have adopted this approach.  In the second
approach, we have fitted the
Ly$\alpha$ forest with the higher order lines of Lyman series
(predominantly Ly$\beta$) at wavelengths down to the observational
limit, 3050 \AA\/, for two QSOs at 
$z_{\mathrm{em}} \sim 2.4$ among 8 QSOs presented in this study. This
analysis includes the Ly$\alpha$ forest from lower $z$ in
the regions where these higher order lines are present.
In Section 2, we describe the UVES observations, data reduction and the
Voigt profile fitting.
In Section 3, we present the analysis of the Ly$\alpha$ forest from the
fitted line parameters, such as the number density.
The Ly$\beta$ forest region is discussed in Section 4.
The conclusions are summarized in Section 5.

\begin{table*}
\caption{Observation log}
\label{tab1}
\begin{tabular}{lccccc}
\hline
\noalign{\smallskip}
QSO & $B^{a}$ & $z_{\mathrm{em}}$ & Wavelength & Exp. time & Observing
Date \\
& & & (\AA\/) & (sec) & \\
\noalign{\smallskip}
\hline
\noalign{\smallskip}
HE0515--4414 & 14.9 & 1.719 &
3050--3860 & 19000 & Dec. 14, 18, 1999 \\
Q1101--264 & 16.0 & 2.145 & 3050--3870 & 23400 &
Feb. 10--16, 2000 \\
J2233--606 & 17.5 & 2.238 & 3050--3860 & 16200 &
Oct. 8-12, 1999  \\
& & &3770--4980 & 12300 & Oct. 10-16, 1999  \\
HE1122--1648 & 17.7 & 2.400 & 3050--3870 & 26400 &
Feb. 10--16, 2000 \\
& & & 3760--4975 & 27000 & Feb. 10--16, 2000 \\
HE2217--2818 & 16.0 & 2.413& 3050--3860 & 16200 &
Oct. 5--6, 1999 \\
& & & 3288--4522 & 10800 & Sep. 27--28, 1999  \\
HE1347--2457 & 16.8 & 2.617 & 3760--4975 & 18000
& Feb. 10--16, 2000 \\
Q0302--003 & 18.4 & 3.281 & 4806--5771 & 20000
& Oct. 12--16, 1999 \\
Q0055--269 & 17.9 & 3.655 & 4634--5600 & 18100 & Sep. 20--22, 2000 \\
& & & 4790--5729 & 17000 & Sep. 20--22, 2000 \\
\noalign{\smallskip}
\hline
\end{tabular}
\begin{list}{}{}
\item[$^{\mathrm{a}}$] Taken from the SIMBAD astronomical database.
The magnitude of
HE1347--2457 is from NED.
\end{list}
\end{table*}

\begin{table*}
\caption[]{Analyzed QSOs}
\label{tab2}
\begin{tabular}{lcccccc}
\hline
\noalign{\smallskip}
QSO & $\lambda\lambda$ & $z_{\mathrm{Ly\alpha}}$ &
$dX^{\mathrm{a}}$ & \# of lines$^{\mathrm{b}}$ &
$\overline{\tau}_\ion{H}{i}$ & $<\!z\!>$ \\
\noalign{\smallskip}
\hline
\multicolumn{7}{c}{Sample A}\\
\hline
\noalign{\smallskip}
HE0515--4414 & 3080--3270 & 1.53--1.69 & 0.408 &
63 & 0.086$^{0.049}_{-0.051}$ & 2.1\\
Q1101--264$^{\mathrm{c}}$ & 3230--3400 & 1.66--1.80 &
0.381 & 62 & 0.085$^{0.049}_{-0.051}$ & \\
&  3500--3778 & 1.88--2.08 &
0.685 & 99 & 0.106$^{0.049}_{-0.051}$ & \\
& 3230--3400, 3500--3778 & 1.66--2.08 & 1.066 & 161 &
0.096$^{0.049}_{-0.051}$ & 2.1\\
J2233--606$^{\mathrm{d}}$ &  3400--3890 & 1.80--2.20 &
1.209 & 166 & 0.156$^{0.049}_{-0.051}$ & 2.1\\
HE1122--1648 & 3500--4091 & 1.88--2.37 & 1.518 & 234 &
0.146$^{0.049}_{-0.051}$ & 2.1\\
HE2217--2818  &  3510--4100 & 1.89--2.37 & 1.519 & 214
& 0.130$^{0.049}_{-0.051}$ & 2.1\\
HE1347--2457 &  3760--4335 &
2.09--2.57 &1.575 & 233 & 0.149$^{0.049}_{-0.051}$ & 2.1\\
Q0302--003 &  4808--5150 &
2.96--3.24 & 1.152 & 167 & 0.334$^{0.049}_{-0.051}$ & 3.3\\
Q0055--269 &  4852--5598 &
2.99--3.60 & 2.638 & 419 & 0.421$^{0.049}_{-0.051}$ & 3.3\\
Q0000--263$^{\mathrm{e}}$ &  5450--6100 &
3.48--4.02 & 2.540 & 312 & 0.733$^{0.049}_{-0.051}$ & 3.8\\
\noalign{\smallskip}
\hline
\noalign{\smallskip}
\multicolumn{6}{c}{Sample B}\\
\hline
\noalign{\smallskip}
HE1122--1648 & 3200--3500$^{\mathrm{f}}$ & 1.63--1.88 & 0.680 &  &
0.084$^{0.049}_{-0.051}$ \\
& 3100--3500 & 1.55--1.88 & 0.893 & 21$^{\mathrm{g}}$ & & \\
HE2217--2818  & 3200--3510$^{\mathrm{f}}$ & 1.63--1.89 & 0.704 &
& 0.142$^{0.049}_{-0.051}$ \\
& 3100--3510 & 1.63--1.89 & 0.917 & 34$^{\mathrm{g}}$ & & \\
\noalign{\smallskip}
\hline
\noalign{\smallskip}
\multicolumn{6}{c}{The Ly$\beta$ forest}\\
\hline
\noalign{\smallskip}
HE1122--1648 & 3674--4091 & 2.02--2.37 & 1.09 & 195 & & \\
HE2217--2818  &  3674--4100 & 2.02--2.37 & 1.12 & 190 & & \\
\noalign{\smallskip}
\hline
\end{tabular}
\begin{list}{}{}
\item[$^{\mathrm{a}}$] For $q_{\mathrm{0}}=0$.
\item[$^{\mathrm{b}}$] Over the column density range,
$N_{\ion{H}{i}} = 10^{12.5-17} \ {\mathrm{cm}}^{-2}$.
\item[$^{\mathrm{c}}$] There is a damped Ly$\alpha$ system at $z$ =
1.8386. The Ly$\alpha$ forest regions closer to the damped Ly$\alpha$
system by less than 50 \AA\/ at each side have been excluded in the
study.
\item[$^{\mathrm{d}}$] See also Cristiani \& D'Odorico (2000).
\item[$^{\mathrm{e}}$] Lu et al. (1996).
\item[$^{\mathrm{f}}$] For the spectral region  3100--3200 \AA\/, the
UVES spectra show a sharp decrease in S/N, so the continuum level is
difficult to determine there. Consequently we exclude this region from
the mean optical depth determination. Line counting applies for higher
$\ion{H}{i}$ column densities, where such uncertainties are less
important.
\item[$^{\mathrm{g}}$] For $N_\ion{H}{i} = 10^{13.64-17} \
\mathrm{cm}^{-2}$.
\end{list}
\end{table*}

\section[]{Observations and Data Reductions}

Table~\ref{tab1} lists the observation log for the QSOs observed with
the VLT/UVES.  In addition to HE0515--4414, J2233--606 and HE2217--2818
from KCD, we have included five more QSOs. Note that we extended the
wavelength coverage for the Ly$\alpha$ forest towards J2233--606 and
HE2217--2818 in this study compared to those from KCD.  The spectra
were reduced with {\small MIDAS/UVES} and the resolution is about $R
\sim 45\,000$.  The S/N varies across the spectrum and the typical S/N
in the Ly$\alpha$ forest is $\sim$ 40--50 for all the QSOs except for
HE1347--2457, which is somewhat lower, $\sim 35$.  The spectra were
normalized locally with the 5th and 7th-order polynomial fitting. In
order to avoid the proximity effect, we only include the Ly$\alpha$
forest $3\,000$ km s$^{-1}$ shortward the Ly$\alpha$ emission in this
study.  See KCD for the details of the data reduction.

Traditionally, the Ly$\alpha$ forest has been thought of as originating
in discrete clouds and has thus been analyzed as a collection of {\it
individual lines}. These absorption lines are generally fitted by Voigt
profiles. From the Voigt profile fits, three line parameters (the
absorption redshift, $z$, the \ion{H}{i} column density,
$N_\ion{H}{i}$ in cm$^{-2}$, and the Doppler parameters, 
$b$ in km s$^{-1}$) are derived for each
cloud.  We have used the VPFIT program (Carswell et al.:
http://www.ast.cam.ac.uk/$\sim$rfc/vpfit.html) to fit the lines, and
for blended systems we have added the minimum number of component
clouds to ensure that the reduced $\chi^{2}$ is below an adopted
threshold value of $1.3$.
Voigt profile fitting is not unique (cf. Kirkman \& Tytler 1997; KCD),
but we have fitted all the absorption lines in a consistent manner here
to allow comparisons of the properties between different redshifts and
sight-lines.

We have adopted two approaches to define a sample of line parameters
from VPFIT.
Most previous studies have analysed the Ly$\alpha$ forest longward the
Ly$\beta$ emission line to avoid the confusion with the Ly$\beta$
forest from higher redshift absorbers. Thus, for 
comparison with other studies, we have
used only the Ly$\alpha$ absorption lines longward the Ly$\beta$ emission,
in our first approach.
All the QSOs in
Table~\ref{tab1} have been fitted in this way, without considering
their Ly$\beta$ absorption profiles. The line parameters fitted with
only the Ly$\alpha$ profiles define Sample A.  We restrict our analysis
to systems with $N_\ion{H}{i} \ge 10^{12.5} \mathrm{cm}^{-2}$, the
greatest value for the detection limit for all the quasars studied. 
At $N_\ion{H}{i} = 10^{12.5} \ {\mathrm{cm}}^{-2}$, the
$b$ values in general span from 15 to 45 km s$^{-2}$ 
where the larger value is effectively set by the 
detection limit. In
addition, we have excluded the absorption lines with $N_\ion{H}{i} \ge
10^{17} \mathrm{cm}^{-2}$ in order to avoid Lyman limit systems (Note
that KCD analyses the forest at $N_\ion{H}{i} = 10^{12.5-16}
\mathrm{cm}^{-2}$. Since there are very few lines with $N_\ion{H}{i} =
10^{16-17} \mathrm{cm}^{-2}$, their results would be very similar even
if the $N_\ion{H}{i}$ range were changed to $N_\ion{H}{i} =
10^{12.5-17} \mathrm{cm}^{-2}$).

Note that there is one high column density system with $N_\ion{H}{i} =
10^{17.46} \ \mathrm{cm}^{-2}$ at $z = 3.192$ towards Q0055--269 from
the Ly$\alpha$ absorption profile. There is, however, no corresponding
Lyman limit, indicating that this high-$N_\ion{H}{i}$ value is not
real. In fact, the Ly$\beta$ and the Ly$\gamma$ absorption profiles
corresponding to this system show a complex of at least 2 
lower-$N_\ion{H}{i}$ clouds, which is difficult to de-blend due to severe
line blending from lower-$z$ Ly$\alpha$ forest. However, since we
define Sample A from the results of fitting Ly$\alpha$ absorption
profiles only, we treat this system as one with high $N_\ion{H}{i}$ and
so have omitted it from the sample.
 
Note also that for the optical depth analysis
we have used all the available wavelength ranges between the
Ly$\alpha$ and Ly$\beta$ emission lines without excluding any
high-$N_\ion{H}{i}$ regions, except for
Q1101--264. There is a damped Ly$\alpha$ system at $z=1.8386$ towards
this QSO, and in this case we exclude the regions within 50 \AA\/ of
the central wavelength of this feature.

In the another second approach, we have used the entire spectrum down to
3050\AA\/, fitting the Ly$\alpha$ lines with the higher orders of Lyman
series and adding the lower-$z$ Ly$\alpha$ lines simultaneously.
The S/N at the Ly$\beta$ forest
is usually much lower than that at the Ly$\alpha$ forest and the
continuum fitting becomes more uncertain. It is of not useful to
include the {\it weaker} Ly$\alpha$ forest lines shortward the
Ly$\beta$ emission since its lower S/N would degrade the results from
the statistical analysis. It is of importance, however, to include the
corresponding Ly$\beta$ lines to determine $N_\ion{H}{i}$ and $b$ of
saturated Ly$\alpha$ lines at $N_\ion{H}{i} \ge 10^{14.5}
\ \mathrm{cm}^{-2}$ more reliably. Some of saturated lines are found to
break into several lower-$N_\ion{H}{i}$ lines. Including these {\it
newly recognized }  lines of the Ly$\alpha$ forest at $N_\ion{H}{i} \ge
10^{14.5} \ \mathrm{cm}^{-2}$ is important if we are considering the
absorbers as individual entities for line counting and clustering
analyses.

During this process, it was found that sometimes the continuum needed
to be re-adjusted by small amounts to obtain a satisfactory fit for all
Lyman series including all the identified metal lines.  
Since fitting
the Ly$\beta$ forest region requires a higher S/N to determine a
reliable continuum, not all the QSOs in Table~\ref{tab1} are suitable
for this fitting method. In addition, the rapidly increasing numbers of
the forest lines with $z$ makes it difficult to fit all the spectral
regions simultaneously at $z > 3$. For these reasons, we have selected
two QSOs, HE1122--1648 and HE2217--2818, out of 8 QSOs in the sample.
Both QSOs have S/N of 30--40 at 3200--3500 \AA\/ and of 10--30 at
3100--3200 \AA\/, suitable to explore the Ly$\beta$ forest at $2 < z < 2.4$.

In addition to the Ly$\beta$ forest, this second fitting approach
provides the Ly$\alpha$ forest at lower $z$, i.e. $z \le
(1+z_{\mathrm{em}}) \times (1025.72/1215.67) -1$. Since the Ly$\alpha$
forest in the Ly$\beta$ regions have a lower S/N, we have restricted
this Ly$\alpha$ forest in the Ly$\beta$ forest region only for the
study of higher $N_\ion{H}{i}$ forest at $N_\ion{H}{i} \ge 10^{13}
\ \mathrm{cm}^{-2}$, i.e. the line number density and the mean
\ion{H}{i} opacity, in order to increase our statistics on $z  < 2.5$.
This Ly$\alpha$ forest in the Ly$\beta$ regions defines Sample B in this
study.  

Keep in mind that all the analyses of the Ly$\alpha$ forest here are
from Sample A only, except the line number density and the mean
\ion{H}{i} opacity.  In any figures in subsequent sections filled
circles represent Sample A, while filled squares are for Sample B.
Table~\ref{tab2} lists  the QSOs defined Sample A, Sample B and the
Ly$\beta$ forest.

Metal lines were identified and removed as described in KCD.  Different
transitions of identified metal lines were fully taken into account
when fitting the Ly$\alpha$ forest.  Metal lines could be assumed to be
almost fully identified at $z < 2.5$, especially towards HE1122--1648
and HE2217--2818 where the analyses cover the full wavelength range
down to 3050 \AA\/. The only possible exception is for HE1347--2457,
where incomplete coverage of the Ly$\alpha$ forest may have resulted in
some heavy element lines being missed. At higher-$z$, the
identification of metal lines becomes more problematic due to severe
line blending. Metal contaminations, however, should be less than 5 per
cent at all redshifts.

\begin{figure*}
\begin{center}
\psfig{file=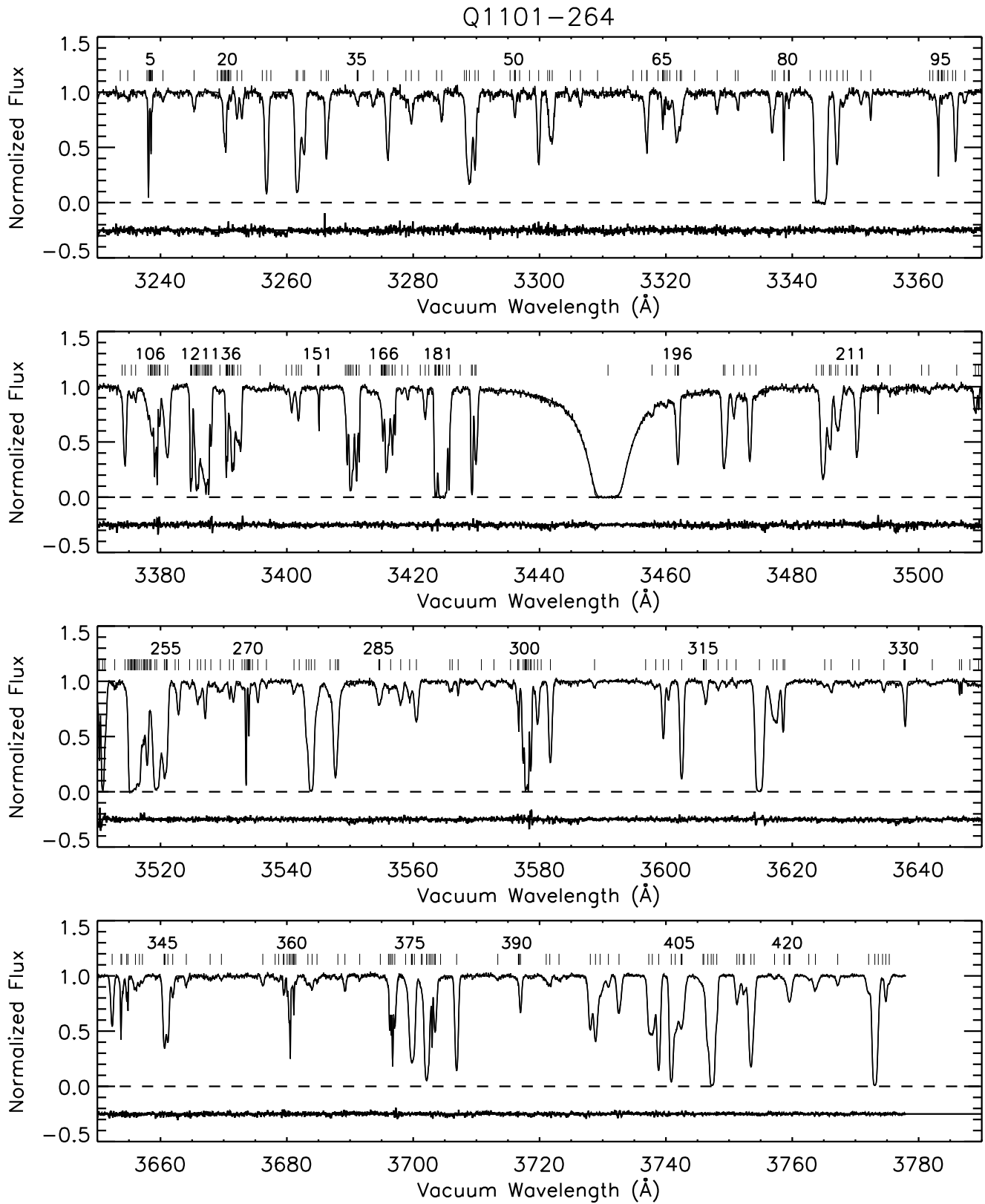}
\end{center}
\caption{The spectrum of Q1101--264 superposed with the
fitted spectrum from the Voigt profile fitting.
The residuals (the differences between the observed
and the fitted flux) shown in the bottom part of each panel are shifted
by $-0.25$.}
\label{fig_q1101_norm1}
\end{figure*}

\begin{figure*}
\begin{center}
\psfig{file=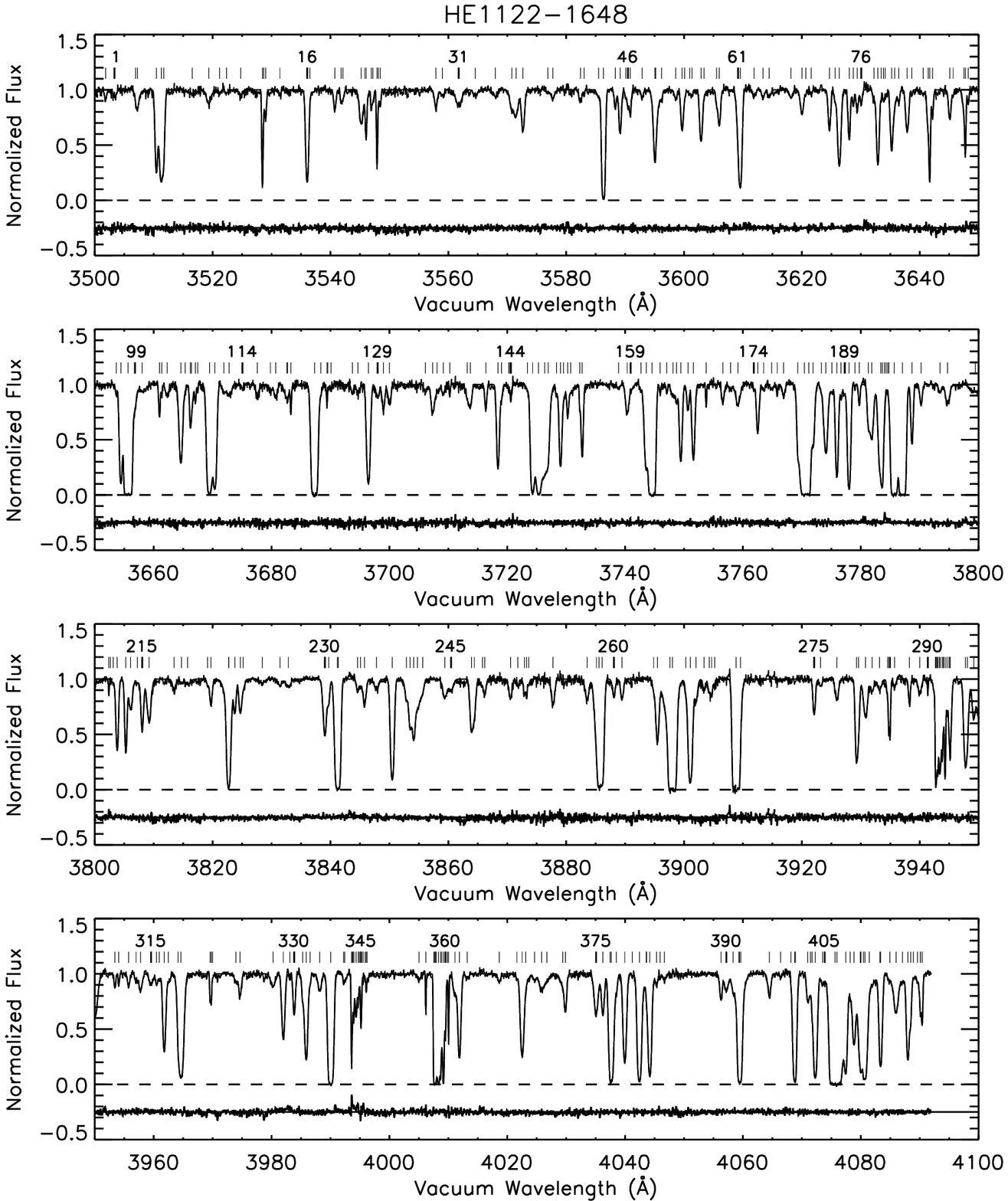}
\end{center}
\caption{The spectrum of HE1122-1648 superposed with the
fitted spectrum from the Voigt profile fitting.
The residuals (the differences between the observed
and the fitted flux) shown in the bottom part of each panel are shifted
by $-0.25$.}
\label{fig_q1122_norm1}
\end{figure*}

\begin{figure*}
\begin{center}
\psfig{file=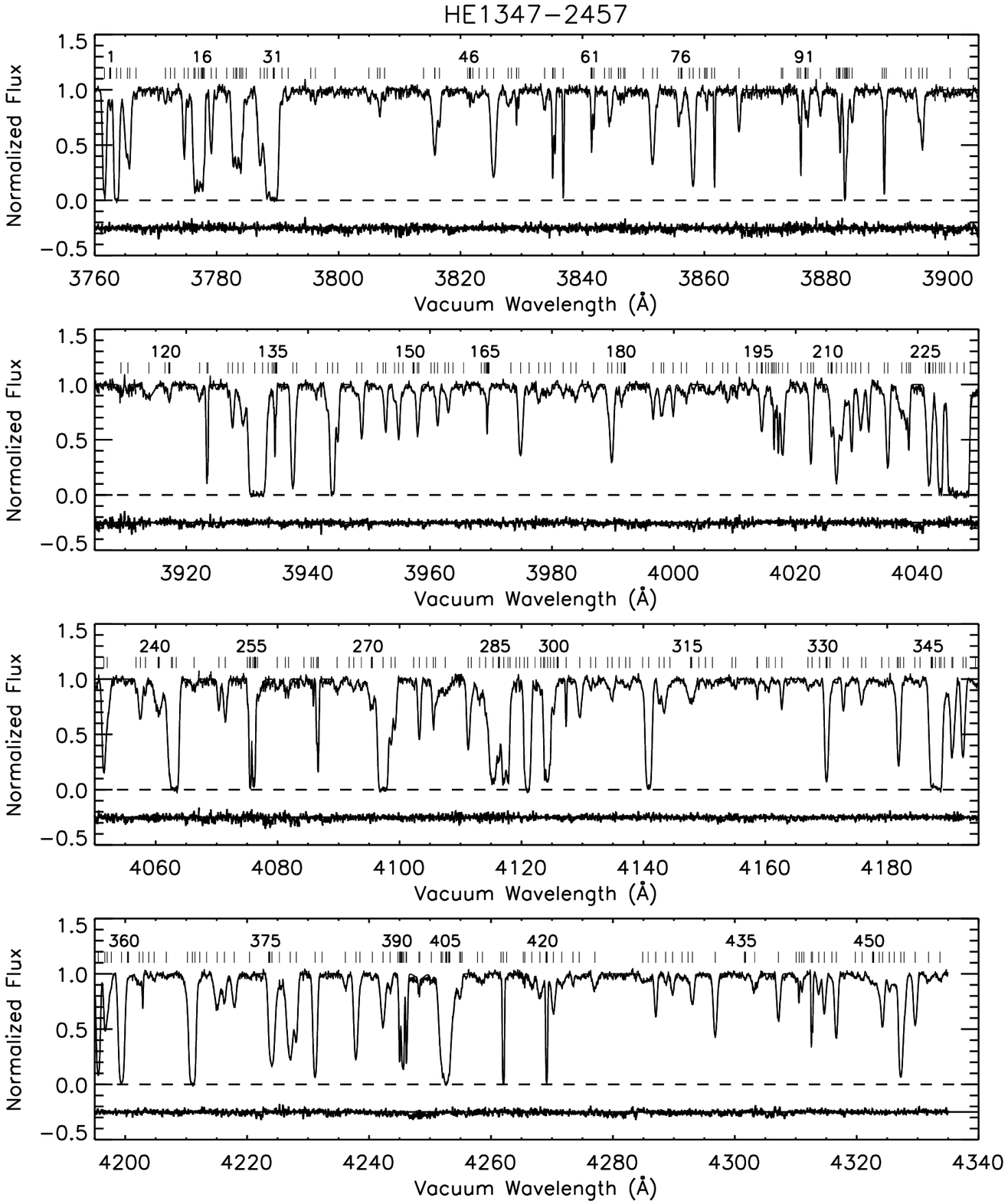}
\end{center}
\caption{The spectrum of HE1347--2457 superposed with the
fitted spectrum from the Voigt profile fitting.
The residuals (the differences between the observed
and the fitted flux) shown in the bottom part of each panel are shifted
by $-0.25$.}
\label{fig_q1347_norm1}
\end{figure*}

\begin{figure*}
\begin{center}
\psfig{file=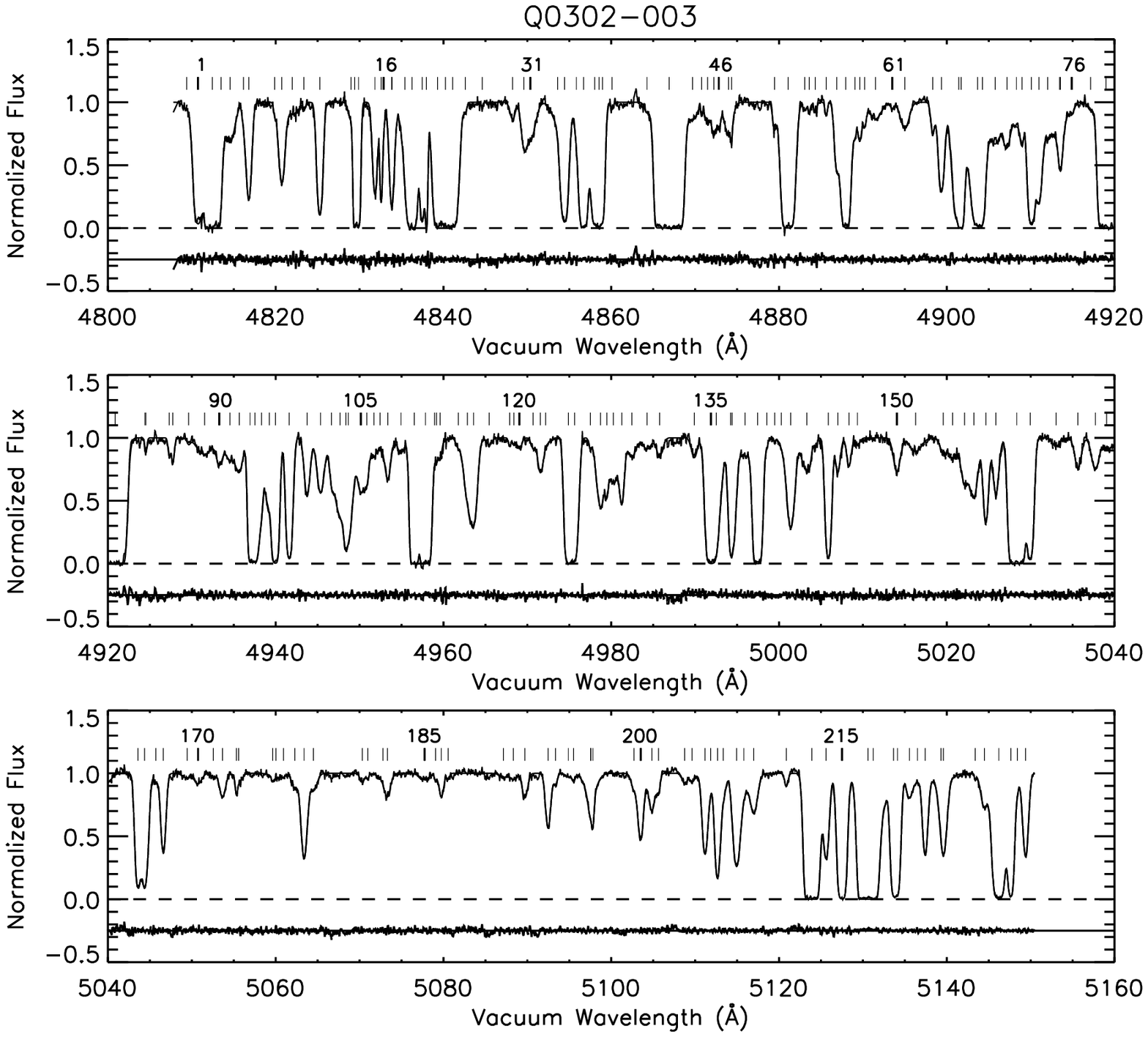}
\end{center}
\vspace{-5.4cm}
\caption{The spectrum of Q0302--003 superposed with the
fitted spectrum from the Voigt profile fitting.
The residuals (the differences between the observed
and the fitted flux) shown in the bottom part of each panel are shifted
by $-0.25$.}
\label{fig_q0302_norm1}
\end{figure*}

Figs.~\ref{fig_q1101_norm1}, \ref{fig_q1122_norm1}, 
\ref{fig_q1347_norm1}, \ref{fig_q0302_norm1} and
\ref{fig_q0055_norm1}
show the Ly$\alpha$ forest used in Sample A: 
Q1101--264, HE1122--1648, HE1347--2457, Q0302--003 and
Q0055--269, respectively. The spectra are
superposed with the fitted spectrum from the Voigt profile
analysis (the sample A fitted line lists from the Voigt profile analysis with
their errors are available electronically at the CDS via anonymous ftp
to cdsarc.u-strasbg.fr (130.79.128.5). An example of the line lists
is shown in Appendix in case of Q1101--264. The line lists of 
HE1122--1648 and HE2217-2818,
including the Ly$\beta$ and Ly$\gamma$ lines down to 3050\AA\/, i.e.
Sample B, will be published elsewhere).
The tick marks indicate the center of
the lines fitted with VPFIT and the numbers above the bold tick
marks indicate the number of the fitted line in the line lists.

\begin{figure*}
\begin{center}
\psfig{file=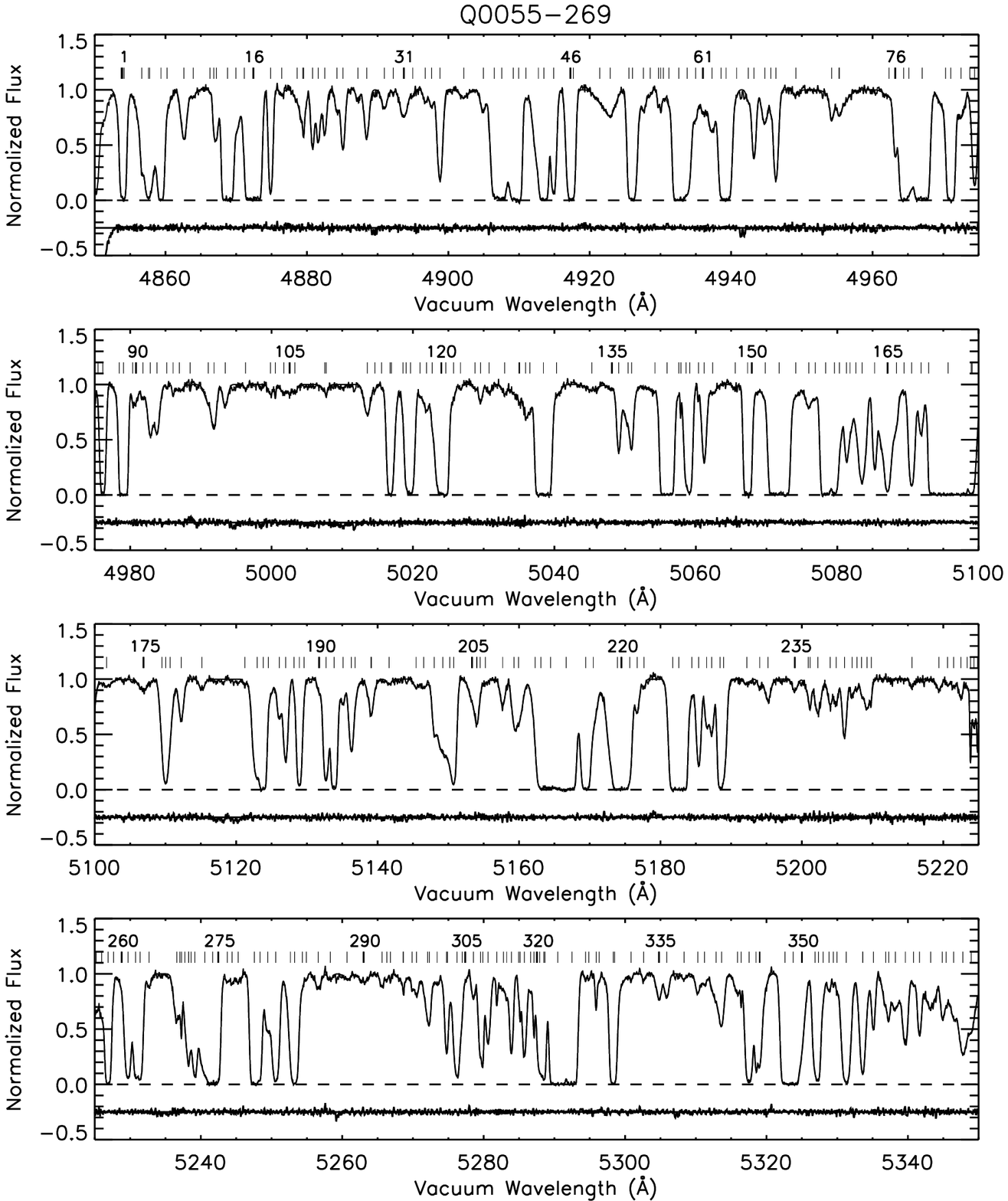}
\end{center}
\caption{The spectrum of Q0055--269 superposed with the
fitted spectrum from the Voigt profile fitting.
The residuals (the differences between the observed
and the fitted flux) shown in the bottom part of each panel are shifted
by $-0.25$.}
\label{fig_q0055_norm1}
\end{figure*}

\setcounter{figure}{5}
\begin{figure*}
\begin{center}
\psfig{file=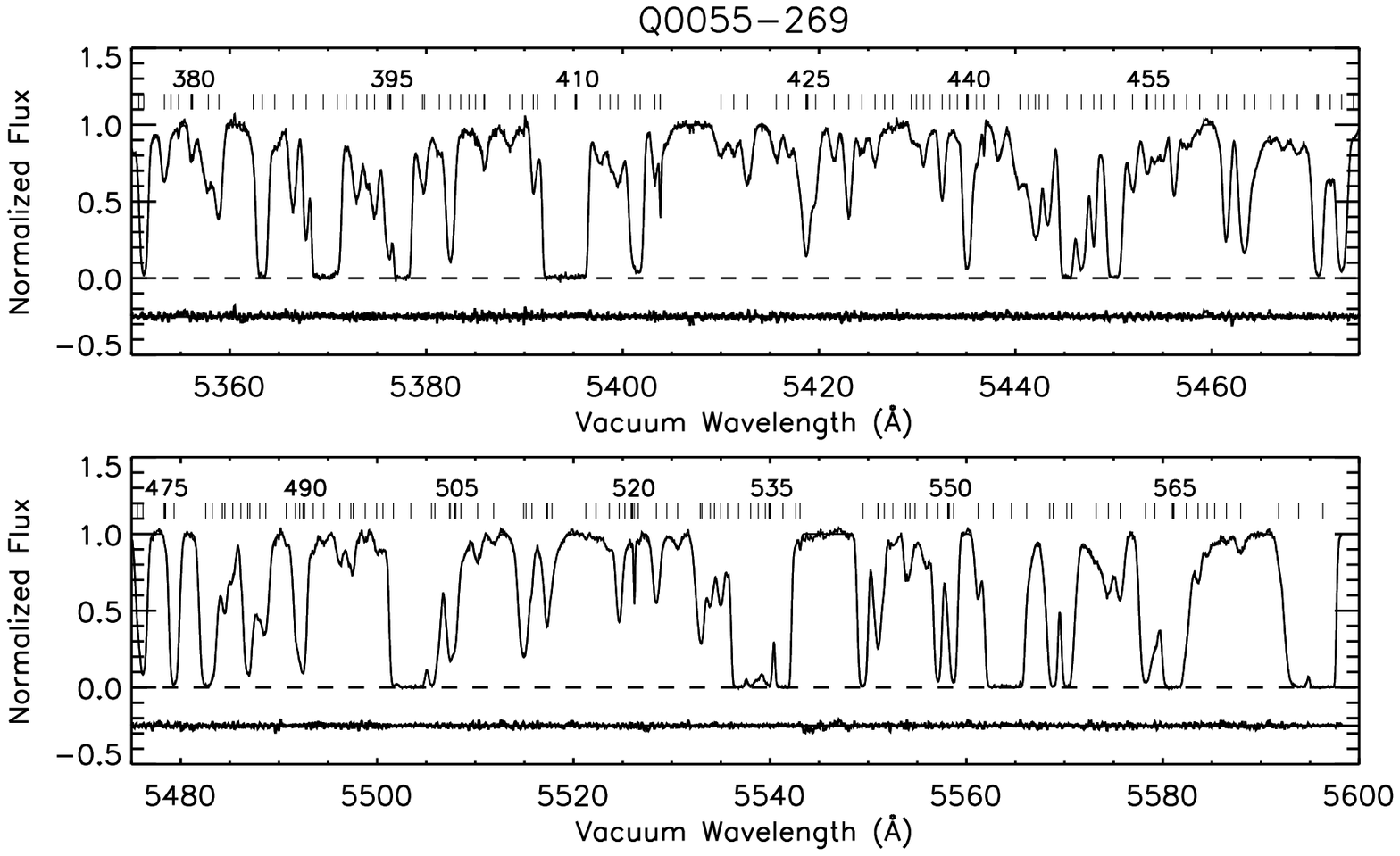}
\end{center}
\vspace{-10.5cm}
\contcaption{}
\end{figure*}

\section{The Ly$\alpha$ forest}

\subsection{The evolution of the line number density}

The line number density per unit redshift is defined as the number of
the forest above a given $N_\ion{H}{i}$ per unit redshift. It is
empirically described by $dn/dz = (dn/dz)_{\mathrm{0}} (1+z)^{\gamma}$,
where $(dn/dz)_{\mathrm{0}}$ is the local comoving number density of
the forest.  For a non-evolving forest in the standard Friedmann
universe with the cosmological constant $\Lambda = 0$ and the constant
UV background, $\gamma = 1$ and 0.5 for $q_{\mathrm{0}}=0$ and 0.5,
respectively. Note that the measured $\gamma$ is dependent on the
chosen column density thresholds, the redshift ranges and the spectral
resolution (Kim et al. 1997; KCD).

\begin{figure}
\begin{center}
\psfig{file=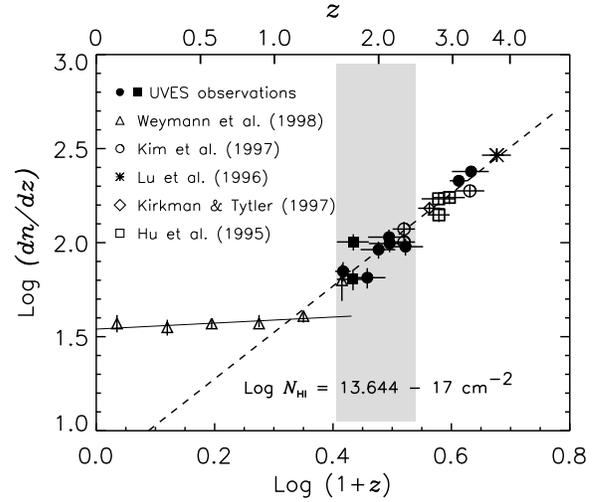, width=8.65cm}
\end{center}
\caption{
The number density evolution of the Ly$\alpha$ forest
over the
column density range
$N_\ion{H}{i} = 10^{13.64 - 17} \ {\mathrm{cm}}^{-2}$, which is
comparable to the
{\it HST} data (open triangles) of Weymann et al. (1998).
The data are shown for the binned sample for display.
The filled symbols are derived from the UVES observations.
Open squares, the star, open circles, and the diamond are taken
from the HIRES data by Hu et al. (1995), Lu et al. (1996),
Kim et al. (1997), 
and Kirkman \& Tytler (1997), respectively.
The horizontal error bars represent the $z$ interval over which
the number density was estimated. The vertical error bars represent
the Poisson $1\sigma$ error.}
\label{fig_dndz1}
\end{figure}

Fig.~\ref{fig_dndz1} shows the number density evolution of the
Ly$\alpha$ forest in the interval $N_\ion{H}{i} = 10^{13.64 - 17}
\ {\mathrm{cm}}^{-2}$ from Sample A (filled circles) and Sample B
(filled squares). This threshold has been chosen to be comparable to
the equivalent width threshold of 0.24 \AA\/ from the {\it HST} QSO
absorption line key project (Weymann et al. 1998),
assuming $N_\ion{H}{i} = 1.33
\times 10^{20} W/\lambda_{\mathrm{0}}^{2}f$, where $W$ is the
equivalent width in angstrom, $\lambda_{\mathrm{0}}$ is the wavelength
of Ly$\alpha$ in angstrom, and $f$ is the oscillator strength of
Ly$\alpha$.

The dashed line represents the maximum-likelihood fit to the UVES and
the HIRES data at $z > 1.5$.  The line number density of HE2217--2818
at $z \sim 1.8$ (Sample B) is higher than that of HE1122--1648 at the
same $z$ range. On the other hand, the triangle at $z \sim 1.55$ from
the {\it HST} observation of UM18 (which is considered as an outlier by
Weymann et al. 1998) is well fit to the power-law derived at $z >
1.5$.  As pointed out by KCD, $(dn/dz)$ decreases as $z$ decreases at
$1.5 < z < 4$ with a consistent pattern and the change in $dn/dz$
occurs at $z < 1.5$.

The lower resolution ($R \sim 1,000$) of the {\it HST/FOS} data, however,
makes it difficult to compare the results by Weymann et al. (1998) to
the results from higher resolution ground-based observations.  In
addition, it should also be noted that there is no reliable conversion
from the equivalent width, $W$, to $N_\ion{H}{i}$, without knowing the
Doppler parameter, $b$, of the absorption lines. The conversion law
between $W$ and $N_\ion{H}{i}$ used in Fig.~\ref{fig_dndz1} is correct
only if $\tau_\ion{H}{i} < 1$. Due to the low resolution and to the
large uncertainty in the continuum fitting, many absorption lines from
the {\it HST} QSO absorption line key project are unresolved and so
$N_\ion{H}{i}$ may be underestimated.
In the literature, $b$ of 25 km s$^{-1}$ is often assumed,
under which the 0.24 \AA\/ threshold corresponds to $N_\ion{H}{i} =
10^{14} {\mathrm{cm}}^{-2}$ (cf. Savaglio et al. 1999; Penton, Shull \&
Stocke 2000; but see also Dav\'e \& Tripp 2001).

\begin{table}
\caption{The line number density evolution}
\label{tab3}
\begin{tabular}{cccc}
\hline
\noalign{\smallskip}
$N_\ion{H}{i}$ range & Median $N_\ion{H}{i}$ &
$(dn/dz)_{\mathrm{0}} $ & $\gamma$ \\
$({\mathrm{cm}}^{-2})$ & $({\mathrm{cm}}^{-2})$ &  & \\
\noalign{\smallskip}
\hline
\noalign{\smallskip}
$10^{13.1-14,\,{\mathrm{a}}}$ & $10^{13.5}$ & $50.0$ & $1.18 \pm 0.14$ \\ 
$10^{13.1-14,\,{\mathrm{b}}}$ & $10^{13.5}$ & $35.0$ & $1.42 \pm 0.16$ \\
$10^{13.64-17}$ & $10^{14.0}$ & $6.1$ & $2.47 \pm 0.18$ \\
$10^{14-17}$ & $10^{14.3}$ & $1.8$ & $2.90 \pm 0.25$ \\
$10^{14.5-17}$ & $10^{15.0}$ & $0.5$ & $3.11 \pm 0.42$ \\
\noalign{\smallskip}
\hline
\end{tabular}
\begin{list}{}{}
\item[$^{\mathrm{a}}$] All the QSOs included.
\item[$^{\mathrm{b}}$] Sample B of HE2217--2818 and HS1946+7658 are excluded.
\end{list}
\end{table}

\begin{figure}
\begin{center}
\psfig{file=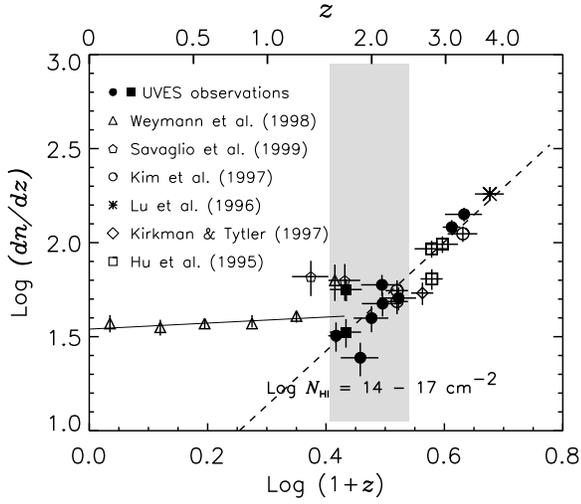, width=8.65cm}
\end{center}
\caption{The number density evolution of the Ly$\alpha$ forest
over the column density range
$N_\ion{H}{i} = 10^{14 - 17} \ {\rm cm}^{-2}$. All the symbols
have the same meaning as in Fig.~\ref{fig_dndz1}.}
\label{fig_dndz2}
\end{figure}

Fig.~\ref{fig_dndz2} is the same as in Fig.~\ref{fig_dndz1}, except the
$N_\ion{H}{i}$ range in $dn/dz$, $N_\ion{H}{i} = 10^{14 - 17}
\ {\mathrm{cm}}^{-2}$.  The dashed line is the maximum-likelihood fit
for the UVES and HIRES observations. 
The points indicated by pentagons in Fig.~\ref{fig_dndz2} are taken
from the Savaglio et al. (1999) analysis of the spectrum of J2233--606
in the Ly$\beta$ and the Ly$\gamma$ regions. There are two important
observational results evident from this figure.  First, the Ly$\alpha$
forest at $1.5 < z < 2$ shows lower $dn/dz$ than at $z < 1.5$, except
the HE2217--2818 forest. Although the HE0515--4414 forest and the
HE1122--1648 forest could be considered to be consistent with the
$dn/dz$ at $z \sim 1$ within 2$\sigma$, the Q1101--264 forest shows
more than 3$\sigma$ difference. Since the universe expands and the
overdensity evolution is rather smooth as a function of $z$ 
for overdensities corresponding to most Ly$\alpha$ forest clouds, it is
difficult to understand the higher $dn/dz$ at $z < 1.5$ in terms of the
overdensity evolution, in particular, a sharp transition at $z \sim
1.5$ shown in Fig.~\ref{fig_dndz2}\footnote{Since the universe expands
and the higher overdensity region collapses to form a structure, the
same $N_\ion{H}{i}$ at different $z$ samples different overdensity.
The overdensity 0 corresponds to $N_\ion{H}{i} \sim 10^{12.7} \
{\mathrm{cm}}^{-2}$, $N_\ion{H}{i} \sim 10^{13.4}
\ {\mathrm{cm}}^{-2}$, and $N_\ion{H}{i} \sim 10^{14}
\ {\mathrm{cm}}^{-2}$ at $z =$ 2, 3, and 4, respectively (Schaye
2001).}.  Instead, it is likely to be caused by the incorrect
conversion between $W$ and $N_\ion{H}{i}$ as well as the effect from
the different resolutions. In reality, $dn/dz$ 
should be somewhere
in between Fig.~\ref{fig_dndz1} and Fig.~\ref{fig_dndz2}.

Second, the number density of the HE2217--2818 forest at $z \sim 1.8$
(Sample B) corresponds to those of UM18 and J2233--606 at similar $z$.
Even though the $dn/dz$ of the UM18 forest is discarded due to lower
resolution, the HE2217--2818 forest agrees well with the J2233--606
forest at $z \sim 1.8$. In fact, the higher $dn/dz$ from HE2217--2818
and J2233--606 is because that their sightlines include several 
higher-$N_\ion{H}{i}$ systems at $N_\ion{H}{i} \ge 10^{14.5}
\ {\mathrm{cm}}^{-2}$.  This indicates that the lines of sight at $z
<2.5$ might not be as homogeneous as at $z > 2.5$.

\begin{figure}
\begin{center}
\psfig{file=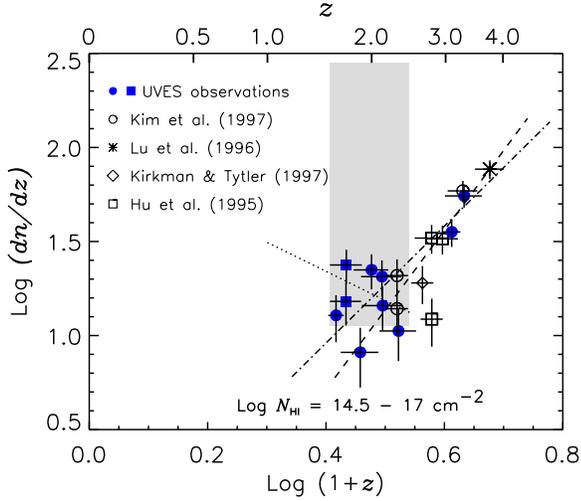, width=8.65cm}
\end{center}
\caption{The number density evolution of the Ly$\alpha$ forest
over the column density range
$N_\ion{H}{i} = 10^{14.5 - 17} \ {\rm cm}^{-2}$. All the symbols
have the same meaning as in Fig.~\ref{fig_dndz1}. The dot-dashed
line represents the single maximum likelihoold fit over the entire
redshift ranges (see Table~\ref{tab3}).
The dotted line indicates the maximum likelihood fit for $z < 2.5$:
$dn/dz = 91.3 \, (1+z)^{-1.55 \pm 1.49}$. The dashed line represents the
maximum likelihood fit for $z > 2.5$: $dn/dz = 0.1 \, (1+z)^{4.24
\pm 0.88}$. The redshift 2.5 is chosen arbitrary.}
\label{fig_dndz3}
\end{figure}

In Fig.~\ref{fig_dndz3}, the dashed line is the
maximum likelihood power law fit for $z > 2.5$ which has been extrapolated
to lower redshifts.
The dotted line gives a maximum likelihood fit for $z < 2.5$.
A single power law does not give a satisfactory fit over
the entire range (the dot-dashed line). 
Because of the small number of
sightlines and the different $dz$ ranges for each sightline,
it is not obvious how to interpret the $dn/dz$ vs $z$
results for this column density range. The general trend may be
for $dn/dz$ to decrease with decreasing $z$, with inhomogeneity 
accounting for some of the higher $dn/dz$ values at lower redshifts.
Alternatively, $dn/dz$ could be almost constant for $z < 2.5$ with
a large scatter from sightline to sightline.
Theuns et al. (1998) predict that $dn/dz$
for the same column density range would show
a non-evolution at $z < 2$. They do not, however, predict any
spatial variation of $dn/dz$.

When extrapolated from the behaviour of $dn/dz$ at lower column density
range, however, the former explanation gives
a better interpretation of Fig.~\ref{fig_dndz3}. 
In short, the higher
$N_\ion{H}{i}$ systems evolve more rapidly and show more evolved
structures in their distributions along different lines of sight.

\begin{figure}
\begin{center}
\psfig{file=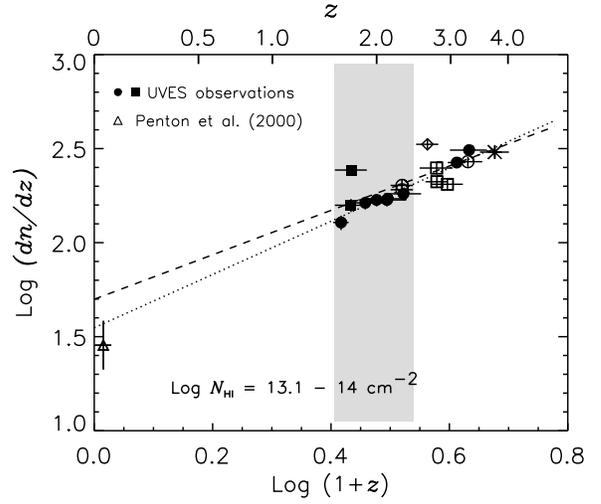, width=8.65cm}
\end{center}
\caption{The number density evolution of the Ly$\alpha$ forest
over the column density range
$N_\ion{H}{i} = 10^{13.1 - 14} \ {\rm cm}^{-2}$.
All the symbols unlisted in the figure are the same as in 
Fig.~\ref{fig_dndz1}.}
\label{fig_dndz4}
\end{figure}

Fig.~\ref{fig_dndz4} is the same as in Fig.~\ref{fig_dndz1}, except the
$N_\ion{H}{i}$ range used is $N_\ion{H}{i} = 10^{13.1 - 14}
\ {\mathrm{cm}}^{-2}$. The dashed line is the maximum-likelihood fit
for the UVES and HIRES observations.
As pointed out by KCD, the $dn/dz$
at $z \sim 2.7$ (the diamond: HS1946+7658) by Kirkman \& Tytler (1997)
is likely due to an over-fitting of higher S/N data than that of the
rest of the observations. In fact, its $\overline{\tau}_\ion{H}{i}$
fits within the fitted power-law (see Fig.~\ref{fig_tau}). On the other
hand, the filled square at $z \sim 1.8$ more than $3\sigma$ above the
$dn/dz$ at similar $z$ is from the HE2217--2818 forest. It also shows a
higher $\overline{\tau}_\ion{H}{i}$ and a higher $dn/dz$ than the
fitted power-law due to several high $N_\ion{H}{i}$ forest.  The dotted
line represents the maximum-likelihood fit when Sample B of the
HE2217--2818 forest and the HS1946+7658 forest are excluded.  

In Fig.~\ref{fig_dndz4}, we assume that line blending at 
this $N_\ion{H}{i}$ range is not severe. Line blending, however,
could lead to underestimate 
the line number density by as much as $\sim 30$ per
cent at $z \sim 3$ (see Hu et al. 1995; Giallongo et al. 1996). This
would cause in part a slower evolution of line number density at $z > 3$.
Although there is a line of sight with a different behavior, the lower
$N_\ion{H}{i}$ forest also evolves with $z$ monotonically
at $1.5 <z < 4$ in terms of the observed line density.

\begin{figure}
\begin{center}
\psfig{file=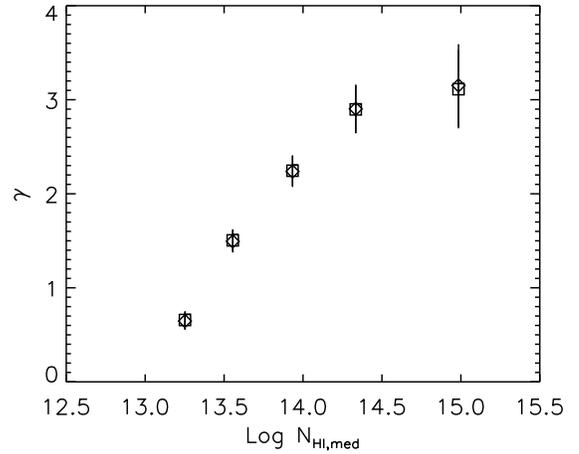, width=8.65cm}
\end{center}
\caption{
The value $\gamma$ as a function of the median $N_\ion{H}{i}$
at the different column density ranges. The upper $N_\ion{H}{i}$ threshold
is fixed to be $N_\ion{H}{i} = 10^{17} \ {\rm cm}^{-2}$
(squares) and $N_\ion{H}{i} = 10^{16} \ {\rm cm}^{-2}$
(diamonds), respectively. The lower $N_\ion{H}{i}$
threshold for each median $N_\ion{H}{i}$, $N_{\rm \ion{H}{i}, med}
= 10^{13.3}, \ 10^{13.6}, \ 10^{13.9}, 
\ 10^{14.3}$, and $10^{15.0} \ {\rm cm}^{-2}$,
corresponds to be $N_\ion{H}{i} =
10^{12.5}, \ 10^{13}, 
\ 10^{13.5}, \ 10^{14}$, and $10^{14.5} \ {\rm cm}^{-2}$, 
respectively.}
\label{fig_gamma}
\end{figure}

Fig.~\ref{fig_gamma} shows $\gamma$ as a function of the median 
$N_\ion{H}{i}$ for the different column density ranges used.
Squares and diamonds represent the 
upper $N_\ion{H}{i}$ threshold as $N_\ion{H}{i}
= 10^{17} \ {\mathrm{cm}}^{-2}$ and $N_\ion{H}{i} = 10^{16}
\ {\mathrm{cm}}^{-2}$, respectively.  Due to a lower number of the
forest at $N_\ion{H}{i} = 10^{16-17} \ {\mathrm{cm}}^{-2}$, $\gamma$
does not change significantly with the upper $N_\ion{H}{i}$ threshold.
The $\gamma$ values increase as the median $N_\ion{H}{i}$ 
increases (Lu et al. 1991; Weymann et al. 1998). This result
could be explained by two different scenarios: 
blending and incompleteness of lower $N_\ion{H}{i}$ at higher $z$, and
a change in the intrinsic properties of the Ly$\alpha$ forest. Keep in
mind that the evolution of the UV background would predict the same
$\gamma$ for all different column density ranges (Dav\'e et al. 1999).

Lower-$N_\ion{H}{I}$
lines are lost at higher $z$ due to more severe line blending, 
i.e. incompleteness, resulting more line loss
at higher $z$ than at lower $z$. As a result, $\gamma$
could become smaller
when lower-$N_\ion{H}{i}$ lines are included in the analysis
(Giallongo et al. 1996; Kim et al. 1997; Dav\'e et al. 1999).
When available lines decrease at $N_{\mathrm{\ion{H}{i}, th}} \ge
10^{14.5} \ {\mathrm{cm}}^{-2}$, $\gamma$ becomes very uncertain.
At $N_{\mathrm{\ion{H}{i}, th}} \sim 10^{13.5-14.5} \ {\mathrm{cm}}^{-2}$,
$\gamma$ shows a value of $\sim 2.3$.

Unfortunately, quantifying the amount of
line blending is not straightforward. The incompleteness corrections considered
in the literature have assumed a single power-law of the column density
distribution. At lower-$N_\ion{H}{i}$, there is a trend of higher
power-law index as $z$ decreases. This could
not only due to incompleteness but also
due to the real structure evolution. At higher-$N_\ion{H}{i}$ at
$N_\ion{H}{i} > 10^{14} \ {\mathrm{cm}}^{-2}$, the onset of rapid,
non-linear collapse results in a deviation from a single power-law
in the column density distribution, which is a function of $z$. 
Extensive Monte-Carlo simulations are required to establish the
level of line blending as a function of $z$ because there are
large uncertainties in the continuum adjustments and varying 
S/N across the observed spectra.
At the same time, these results should be
compared with the ones from numerical simulations in order to constrain
the significance of blending on the $\gamma$--median $N_\ion{H}{i}$
relation.

On the other hand, the increase of $\gamma$ as a function of the
median $N_\ion{H}{i}$ could be a result of the intrinsic evolution
of the Ly$\alpha$ forest itself. Weaker Ly$\alpha$ forest arising
from lower density gas expands faster than stronger Ly$\alpha$ forest.
Thus the fractional cross section to detect weaker Ly$\alpha$ forest
increases as $z$ decreases (Dav\'e et al. 1999). 

It is obvious, however, that a stronger variation in $dn/dz$ for 
higher $N_\ion{H}{i}$ clouds also plays a part in the 
$\gamma$--median $N_\ion{H}{i}$ relation
(see Fig.~\ref{fig_dndz3}). Due to a couple of higher $dn/dz$ at $z < 2.5$
for higher
column density forest, the
exponent from the maximum likelihood fit becomes smaller.

\subsection{The evolutionary behavior of $dn/dz$}

As shown in Figs.~\ref{fig_dndz1}--\ref{fig_dndz4}, the break point in
$dn/dz$ may occur at redshifts as low as $z \sim 1$, and very likely
$z < 1.5$, rather than $z \sim 1.7$--2 as suggested by Theuns et al.
(1998) and Dav\'e et al. (1999).  This shows that the UV background
does not decrease as rapidly as the QSO-dominated UV background by
Haardt \& Madau (1996) used in simulations, and so galaxies could be a
main source of the UV background at $z < 2.5$ and/or that the structure
formation could be underestimated in simulations.  Dav\'e et al. (1999)
show that $dn/dz \propto (1+z)^{5\beta-6.5}$ for a constant UV
background in the flat Friedmann universe. For $N_\ion{H}{i}=10^{13.64-17}
\ {\mathrm{cm}}^{-2}$ at $1.5 < z < 4$, $\beta = 1.74 \pm 0.02$.  
This indicates $dn/dz
\propto (1+z)^{2.25}$, which is in agreement with the observations,
$dn/dz \propto (1+z)^{2.44 \pm 0.18}$.  This result suggests a
non-decreasing UV background at $1.5 < z < 4$, unlike the decreasing 
one expected
from the Haardt \& Madau UV background at $z < 2.5$, i.e. the QSOs alone
might not be enough to explain the $dn/dz$ evolution of the forest
(KCD; Bianchi et al. 2001).

Besides the break point in $dn/dz$ at $z \sim 1$, Lu et al. (1991)
suggest that $dn/dz$ could be fit better with a double power-law at $z
\sim 2.3$ for $1.6 < z < 3$, using the minimum $W$ threshold
$W=0.36$\AA\/. The minimum $W$ threshold corresponds to 
$N_\ion{H}{i}=10^{14.7}, \ 10^{14.4}$ and $10^{14.2} \ {\mathrm{cm}}^{-2}$
for $b=25$, 30 and 35 km s$^{-1}$, respectively. This threshold
is 
similar to the one used in Fig.~\ref{fig_dndz3}. As clearly seen
in Fig.~\ref{fig_dndz3}, there is no clear trend of a different evolution
of $dn/dz$ at higher and lower redshifts.  

The general trend shown in $dn/dz$ confirms that $dn/dz$ continues its
evolution in the same general manner at least for $1.5 < z < 4$, which
indicates the importance of the Hubble expansion over the structure
formation at this redshift ranges (Miralda-Escud\'e et al. 1996; Theuns
et al. 1998; Dav\'e et al. 1999).  Our new observations
(Fig.~\ref{fig_dndz3}), however, show that the structure evolution
becomes more important in some lines of sights at $z < 2.5$ and the
clustering of stronger lines also increases as $z$ decreases
(Fig.~\ref{fig_step}).

\subsection{The mean \ion{H}{i} opacity}

As seen in Section 3.1, comparing the the line number densities at $z >
1.5$ with those at $z < 1.5$ is somewhat uncertain because of the
different spectral resolutions used.  The mean \ion{H}{i} opacity
provides more straightforward comparisons between different qualities
of data. Moreover, the mean \ion{H}{i} opacity does not rely on the
subjective line counting method, although it is more subject to
continuum uncertainties (KCD).

\begin{figure}
\begin{center}
\psfig{file=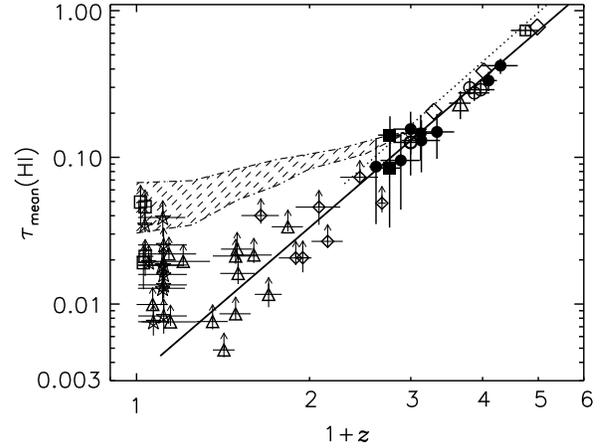, width=8.65cm}
\end{center}
\caption{The \ion{H}{i} opacity as a function of $z$. Filled symbols
represent the mean \ion{H}{i} opacity from the UVES data.  The open
circle at $<\!z\!> \ =2.0$ represents $\overline{\tau}_\ion{H}{i}$ of
J2233--606 when two high column density systems are excluded.  Other
symbols at $z > 1.5$ are from: open circles (Hu et al. 1995), large
square (Lu et al. 1996), the open triangle (Kirkman \& Tytler 1997),
and diamonds (Rauch et al. 1997).  Symbols at $z < 1.5$ with arrows are
from: open diamonds (Impey et al. 1996), open triangles (Weymann et al.
1998), open stars (Impey et al. 1999) and open squares (Penton et al.
2000). Due to the low resolution and low S/N spectra of the {\it HST}
observations, ${\overline{\tau}}$ could be highly underestimated
if the Ly$\alpha$ forest at $N_\ion{H}{i} \le 10^{14} \
{\mathrm{cm}}^{-2}$ contains the bulk of the neutral hydrogen.
$\overline{\tau}_\ion{H}{i}$ estimates, while the y-axis error bars
were estimated from simply changing the adopted continuum by $\pm
5$\%.  The dotted line represents the commonly used formula by Press et
al. (1993), ${\overline{\tau}}_{\mathrm{\ion{H}{i}}} (z) =
0.0037 \, (1+z)^{3.46}$.
The shaded area enclosed with dot-dashed lines indicates the
ranges of ${\overline{\tau}}$ expected from different cosmological
simulations by Dav\'e et al. (1999).}
\label{fig_tau}
\end{figure}

Fig.~\ref{fig_tau} shows the $\overline{\tau}_\ion{H}{i}$ measurements
(the effective optical depth $\tau_{\mathrm{eff}}$;
$\exp^{-\tau_{\mathrm{eff}}} = \ <\!\exp^{-\tau}\!>$, where $<\,\,>$
indicates the mean value averaged over wavelength), together with other
opacity measurements compiled from the literature.  Filled symbols are
the $\overline{\tau}_\ion{H}{i}$ measurements from the UVES data. The
dotted line represents a widely used $\tau_{\mathrm{eff}}$ from
low-resolution data by Press, Rybicki \& Schneider (1993).
Table~\ref{tab2} lists the estimated $\overline{\tau}_\ion{H}{i}$
values from the UVES observations (see KCD for the actual
$\overline{\tau}_\ion{H}{i}$ values at $z > 1.5$ in
Fig.~\ref{fig_tau}, which is not listed in Table~\ref{tab2}). 
Table~\ref{tab4} lists the estimated
$\overline{\tau}_\ion{H}{i}$ values from {\it HST} observations, adding the
equivalent widths. The different $\overline{\tau}_\ion{H}{i}$ values of
3C273 at the different $z$ ranges by different studies suggest the
uncertainty in deriving $\overline{\tau}_\ion{H}{i}$ as well as a
small-scale cosmic variance.

The solid line represents the least-squares fit to the UVES (only
filled circles and squares) and HIRES data:
$\overline{\tau}_\ion{H}{i} (z) = (0.0032 \pm 0.0009) \, (1+z)^{3.37
\pm 0.20}$. As stated in KCD, $\overline{\tau}_\ion{H}{i}$ is well fit
to a single power-law at $1.5 < z < 4$ and the newly fit power-law is
about a factor of 1.3 smaller than the Press et al. value from their
lower resolution data.

The square with higher $\overline{\tau}_\ion{H}{i}$ at $z \sim 1.8$ is
from HE2217--2818, which shows several high-$N_\ion{H}{i}$ clouds at
$N_{\ion{H}{i}} = 10^{14-17} {\mathrm{cm}}^{-2}$.  This increases
$\overline{\tau}_\ion{H}{i}$ at $z \sim 1.8$, compared to that of
HE1122--1648 without high-$N_\ion{H}{i}$ clouds at the same redshift
range. The bold open circle at $z \sim 2$ is
$\overline{\tau}_\ion{H}{i}$ towards J2233--606, when two higher column
density absorption systems were excluded.  Note that
$\overline{\tau}_\ion{H}{i}$ towards HE2217--2818 at $z \sim 1.8$ and
J2233--606 are similar to the Press et al. value at the same
redshifts.  This confirms that the higher Press et al. value is in part
due to the inclusion of high column density systems in their low
resolution sample as well as the uncertainty in the continuum
displacement, as noted by KCD. Keep in mind that most QSOs used in
our analysis both from the UVES and Keck observations do show few
high column density systems at $N_{\ion{H}{i}} > 10^{17} 
{\mathrm{cm}}^{-2}$. In the case of having a damped system 
along the sightline, such as Q1101--264 and Q0000--263, we excluded 
the regions in which the damped systems are located.   

There is a large uncertainty in $\overline{\tau}_\ion{H}{i}$ from {\it HST}
observations due to their lower S/N, lower resolution, and lower
absorption line densities with the presence of weak emission lines, all
resulting in an unreliable local continuum fit.  This, in general,
leads to a tendency to miss weak lines and so an underestimate
$\overline{\tau}_\ion{H}{i}$.  There is, however, indication of
slow-down in the evolution of $\overline{\tau}_\ion{H}{i}$. Despite a
large scatter at $z \sim 0.1$, the median $\overline{\tau}_\ion{H}{i}$
at $z \sim 0.1$ is 0.019, a factor of 4.3 larger than the value
extrapolated from at $z > 1.5$. 
Although open triangles (Weymann et al.
1998) might underestimate $\overline{\tau}_\ion{H}{i}$ significantly,
the different evolution in $\overline{\tau}_\ion{H}{i}$ occurs at $z <
1$, a bit lower than that suggested from $dn/dz$.

Dav\'e et al. (1999) simulated the evolution of 
$\overline{\tau}_\ion{H}{i}$, assuming different cosmologies and
normalizing their results to observations at $z \sim 3$. The shaded
area in Fig.~\ref{fig_tau} shows the ranges of $\overline{\tau}_\ion{H}{i}$
from their simulations at $z < 2$. Although the estimation of 
$\overline{\tau}_\ion{H}{i}$ from {\it HST} observatons could be 
underestimated by a large factor, it is not in good agreement with
the simulated $\overline{\tau}_\ion{H}{i}$. In particular, at $z \sim 0.5$,
there are some observations which show lower $\overline{\tau}_\ion{H}{i}$ 
than the simulated one by more than a factor of 10. 
This could indicate that the QSO-dominated UV background by Haardt \&
Madau (1996) used in the simulations does not represent the {\it real}
UV background at $z < 2$, i.e. this Haardt \& Madau 
QSO-dominated background underestimates
the real UV background at $z < 2$, as suggested by the evolution
of $dn/dz$.
In addition, the structure in the Ly$\alpha$
forest becomes more highly patchy at $z < 2$ than the simulations predict
(see Fig.~\ref{fig_dndz3}). In
the latter case, some lines of sight could have a very lower 
$\overline{\tau}_\ion{H}{i}$.

\begin{table}
\caption{The mean \ion{H}{i} opacity compiled from the literature}
\label{tab4}
\begin{tabular}{ccccc}
\hline
\noalign{\smallskip}
QSO & $z_{\mathrm{em}}$ & $\Delta z$ &
$\overline{\tau}_\ion{H}{i}$ & Ref. \\
\noalign{\smallskip}
\hline
\noalign{\smallskip}
PKS0044+03 & 0.624 &  0.403--0.590 & $0.024^{0.027}_{-0.021}$ &
1\\
3C95 & 0.614 & 0.394--0.580 & $0.009^{0.009}_{-0.008}$ &
1\\
US1867 & 0.513 & 0.357--0.481 & $0.005^{0.005}_{-0.004}$ &
1\\
3C263 & 0.652 & 0.357--0.617 & $0.021^{0.024}_{-0.019}$ &
1\\
3C273 & 0.158 & 0.000--0.134 & $0.010^{0.011}_{-0.009}$ &
1\\
PG1259+593 & 0.472 & 0.020--0.271 & $0.008^{0.008}_{-0.007}$ &
1\\
PG1259+593 & 0.472 & 0.271--0.441 & $0.008^{0.008}_{-0.007}$ &
1\\
3C351 & 0.371 & 0.069--0.342 & $0.020^{0.021}_{-0.018}$ &
1\\
H1821+643 & 0.297 & 0.008--0.270 & $0.022^{0.024}_{-0.020}$ &
1\\
PKS2145+06 & 0.990 & 0.481--0.719 & $0.022^{0.024}_{-0.019}$ &
1\\
PKS2145+06 & 0.990 & 0.719--0.948 & $0.034^{0.036}_{-0.031}$ &
1\\
3C454.3 & 0.859 & 0.398--0.606 & $0.016^{0.019}_{-0.014}$ &
1\\
3C454.3 & 0.859 & 0.606--0.789 & $0.012^{0.014}_{-0.010}$ &
1\\
PG1222+228$^{\mathrm{a}}$ & 2.046 & 0.892--1.262 &
$0.046^{0.057}_{-0.035}$ & 2\\
PG1222+228$^{\mathrm{a}}$ & 2.046 & 1.262--1.631 &
$0.073^{0.080}_{-0.067}$ & 2\\
PG1222+228$^{\mathrm{a}}$ & 2.046 & 1.631--1.715 &
$0.049^{0.055}_{-0.042}$ & 2\\
PG1634+706$^{\mathrm{a}}$ & 1.334 & 0.522--0.769 &
$0.040^{0.045}_{-0.035}$ & 2\\
PG1634+706$^{\mathrm{a}}$ & 1.334 & 0.769--1.016 &
$0.021^{0.023}_{-0.018}$ & 2\\
PG1634+706$^{\mathrm{a}}$ & 1.334 & 1.016--1.285 &
$0.027^{0.029}_{-0.024}$ & 2\\
PG2302+029$^{\mathrm{a}}$ & 1.044 & 0.892--1.001 &
$0.021^{0.025}_{-0.016}$ & 2\\
PG1211+143 & 0.085 & 0.006--0.062 &
$0.035^{0.038}_{-0.031}$ & 3\\
Q1214+1804 & 0.375 & 0.006--0.216 &
$0.039^{0.049}_{-0.029}$ & 3\\
PG1216+069 & 0.334 & 0.006--0.223 &
$0.025^{0.029}_{-0.022}$ & 3\\
PKS1217+023 & 0.240 & 0.006--0.214 &
$0.013^{0.015}_{-0.010}$ & 3\\
3C273 & 0.158 & 0.006--0.134 &
$0.008^{0.009}_{-0.006}$ & 3\\
J1230.8+0115 & 0.117 & 0.007--0.093 &
$0.019^{0.021}_{-0.017}$ & 3\\
Q1228+1116 & 0.235 & 0.007--0.209 &
$0.018^{0.023}_{-0.014}$ & 3\\
Q1230+0947 & 0.420 & 0.007--0.223 &
$0.013^{0.016}_{-0.011}$ & 3\\
Q1245--0333 & 0.379 & 0.007--0.223 &
$0.018^{0.022}_{-0.015}$ & 3\\
PKS1252+119 & 0.870 & 0.007--0.223 &
$0.016^{0.020}_{-0.012}$ & 3\\
Q1252+0200 & 0.345 & 0.007--0.223 &
$0.008^{0.010}_{-0.006}$ & 3\\
3C273 & 0.158 & 0.000--0.070 & $0.021^{0.027}_{-0.016}$ &
4\\
H1821+643 & 0.297 & 0.013--0.042 & $0.019^{0.026}_{-0.013}$ &
4\\
PKS2155--304 & 0.117 & 0.006--0.064 & $0.046^{0.054}_{-0.038}$ &
4\\
Q1230+0115 & 0.117 & 0.001--0.032 & $0.050^{0.064}_{-0.035}$ &
4\\
\noalign{\smallskip}
\hline
\end{tabular}
\begin{list}{}{}
\item[$^{\mathrm{a}}$] The equivalent widths listed in the referenced
paper are only for the core of the profile. Impey et al. (1996)
gave a scale factor of 1.33 for isolated lines and of 1.18 for
blended lines, to recover the total equivalent widths. We have used
a scale factor of 1.255, a mean value.
\item[Ref.:] 1. Bahcall et al. (1993);
2. Impey et al. (1996);
3. Impey et al. (1999);
4. Penton et al. (2000).
\end{list}
\end{table}

\subsection{The differential density distribution function}

The differential density distribution function, $f(N_\ion{H}{i})$, is
defined as the number of the absorption lines per unit absorption
distance path and per unit column density as a function of
$N_\ion{H}{i}$.  The absorption distance path $X(z)$ is defined by
$X(z) \equiv {1\over 2} [(1+z)^{2} -1]$ for $q_{\mathrm{o}} = 0$ or by
$X(z) \equiv {2\over 3} [(1+z)^{3/2} -1]$ for $q_{\mathrm{o}} = 0.5$ in
the standard Friedmann universe (see Table~\ref{tab2} for
$q_{\mathrm{o}} = 0$).  Empirically, $f(N_\ion{H}{i})$ is fitted to a
power law:  $f(N_\ion{H}{i}) = A \, N_\ion{H}{i}^{-\beta}$.

\begin{figure}
\begin{center}
\psfig{file=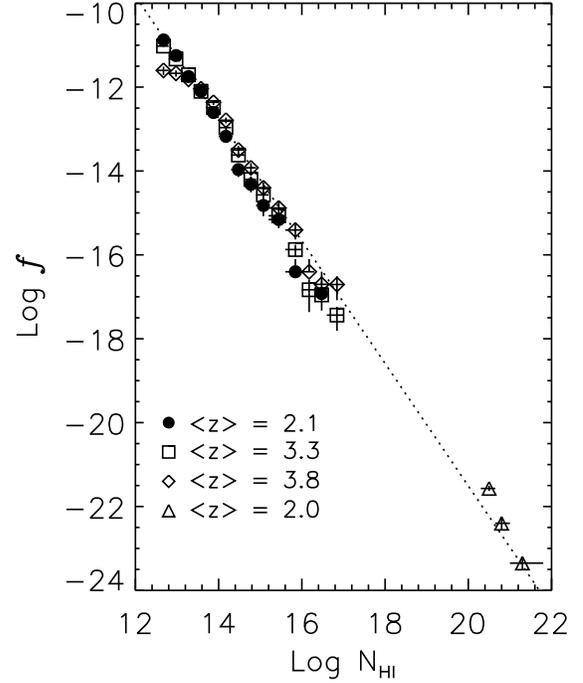, width=8.65cm}
\end{center}
\caption{The differential density distribution functions at \hfill
$<\!z\!> \, = 2.1$, 3.3 and 3.8, without the incompleteness correction.
The data are shown in the binned sample for display.  Open triangles
are the differential density distribution functions for the damped
Ly$\alpha$ systems at $1.5 < z < 2.5$ (Storrie-Lombardi \& Wolfe
2000).  The dotted line represents the incompleteness-corrected
$f(N_\ion{H}{i})$ at $<\!z\!> \ = 2.85$ from Hu et al. (1995),
$f(N_\ion{H}{i}) = 4.9 \times 10^{7} \, N_\ion{H}{i}^{-1.46}$.}
\label{fig_ddf}
\end{figure}

Fig.~\ref{fig_ddf} shows the observed $f(N_\ion{H}{i})$ at different
redshift ranges without the incompleteness correction due to line blending.
The dotted line represents the incompleteness-corrected
$f(N_\ion{H}{i})$ at $<\!z\!> \ = 2.85$ from Hu et al. (1995),
$f(N_\ion{H}{i}) = 4.9 \times 10^{7} \, N_\ion{H}{i}^{-1.46}$.
Triangles for the damped Ly$\alpha$ systems at $1.5 < z < 2.5$ are
taken from Storrie-Lombardi \& Wolfe (2000).  At $z \sim 2.1$,
$f(N_\ion{H}{i})$ at $N_\ion{H}{i} = 10^{12.5 -14.5}
\ {\mathrm{cm}}^{-2}$ is in good agreement with the
incompleteness-corrected $f(N_\ion{H}{i})$ at $z \sim 2.8$.  It is also
true for $f(N_\ion{H}{i})$ at $z \sim 3.3$, although the
goodness-of-the-fit is lower at $N_\ion{H}{i} = 10^{12.5-13}
\ {\mathrm{cm}}^{-2}$. In general, $f(N_\ion{H}{i})$ is well
approximated to a single power-law $f(N_\ion{H}{i}) \propto
N_\ion{H}{i}^{-1.5}$ at $N_\ion{H}{i} = 10^{12.5 -22}
\ {\mathrm{cm}}^{-2}$ (Petitjean et al. 1993).

As noted by Petitjean et al. (1993) and Kim et al. (1997),
$f(N_\ion{H}{i})$ starts to deviate from the empirical power-law at
$N_\ion{H}{i} > 10^{14} \ {\mathrm{cm}}^{-2}$.  The amount of this
deviation increases as $z$ decreases since the higher $N_\ion{H}{i}$
forest disappears more rapidly as $z$ decreases.  In addition, the
deviation $N_\ion{H}{i}$ at which $f(N_\ion{H}{i})$ starts to deviate
decreases as $z$ decreases. At $<\!z\!> \ =$ 3.8, 3.3 and 2.1,
$f(N_\ion{H}{i})$ deviates from the power-law at $N_\ion{H}{i} \sim
10^{16} \ {\mathrm{cm}}^{-2}$, $N_\ion{H}{i} \sim 10^{14.5}
\ {\mathrm{cm}}^{-2}$ and $N_\ion{H}{i} \sim 10^{14.2}
\ {\mathrm{cm}}^{-2}$, respectively.  At $N_\ion{H}{i} > 10^{15.6}
\ {\mathrm{cm}}^{-2}$, the deviation from the single power-law
increases more than $3\sigma$ at $z \sim 2.1$ and more than $2\sigma$
at $z \sim 3.3$.  Table~\ref{tab5} lists $\log A$ and $\beta$ for the
various column density ranges from the maximum-likelihood fit.

\begin{figure}
\begin{center}
\psfig{file=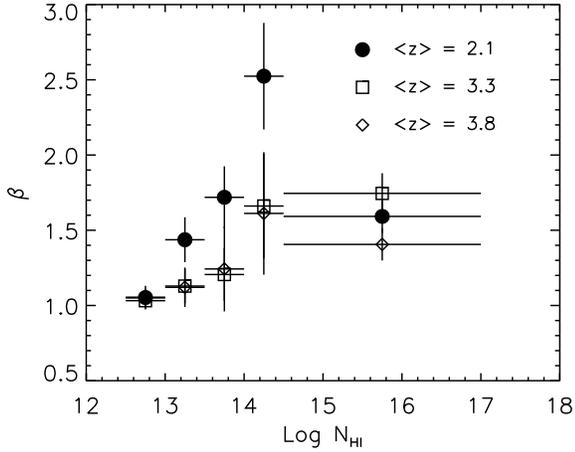, width=8.65cm}
\end{center}
\caption{The slope $\beta$ as a function of the $N_\ion{H}{i}$ range
used in the maximum likelihood fit.}
\label{fig_beta}
\end{figure}

Fig.~\ref{fig_beta} shows $\beta$ as a function of the $N_\ion{H}{i}$
range. The x-axis error bars represent the $N_\ion{H}{i}$ range used in
the fit.  For $N_\ion{H}{i} < 10^{14.5} \ {\mathrm{cm}}^{-2}$, $\beta$
increases as $N_\ion{H}{i}$ increases, in part due to incompleteness at
$N_\ion{H}{i} \le 10^{13} \ {\mathrm{cm}}^{-2}$ (Hu et al. 1995;
Giallongo et al. 1996, see also Section 3.1).  For $N_\ion{H}{i} >
10^{14.5} \ {\mathrm{cm}}^{-2}$, $\beta$ is ill-defined since there are
not enough lines in the samples to get a reliable $\beta$, especially
at $z \sim 2.1$.

Although it is statistically uncertain, there is a suggestion of
increasing $\beta$ as $z$ decreases for a given $N_\ion{H}{i}$ range in
Fig.~\ref{fig_beta}. Even if we exclude the lower-$N_\ion{H}{i}$
range distorted by line blending, the similar trends hold.
Table~\ref{tab5} lists the $\beta$ values at
$<\!z\!> \ = 0.03$ using the conversion from $W$ by Penton et al.
(2000) and the $\beta$ value at $<\!z\!> \ = 0.17$ using the direct
profile fitting analysis by Dav\'e \& Tripp (2001) for different column
density ranges.  Even if we discard the $\beta$ values by Penton et
al. (2000) due to their $W$ analysis, the $\beta$ value at $<\!z\!>
\ = 0.17$ suggests that $\beta$ increases as $z$ decreases.  This is
another way of showing that the strong Ly$\alpha$ forest evolves
rapidly at $0 < z < 4$.

\subsection{The differential mass density distribution of
the Ly$\alpha$ forest}

Schaye (2001) has shown that the shape of the differential density
distribution function reflects that of the differential mass density
distribution function as a function of the gas density.  This
differential mass density distribution function of the Ly$\alpha$
forest is the mass density per unit overdensity $\delta$. 
Assuming that the gas is isothermal,
it can be
described by
\begin{eqnarray*}
\frac{d\Omega_{\mathrm{g}}}{d\log(1+\delta)} \sim
7.6 \times 10^{-9} \, h^{-1} \, 
\Gamma_{12}^{1/3} \, T_{4}^{0.59} \,
\end{eqnarray*}
\begin{equation}
\label{eq2}
\\ \times {\left(\frac{f_{\mathrm{g}}}{0.16}\right)}^{1/3}
\, N_\ion{H}{i}^{4/3}
\, f(N_\ion{H}{i},z),
\end{equation}
\noindent where $N_\ion{H}{i}$ is related to $\delta$
through
\begin{eqnarray*}
N_\ion{H}{i} \sim 2.7 \times 10^{13} (1+\delta)^{1.5}\,
T_{4}^{-0.26} \Gamma^{-1}_{12}
\end{eqnarray*}
\begin{equation}
\label{eq3}
\\ \times {\left(\frac{1+z}{4}\right)}^{9/2} \,
{\left(\frac{\Omega_{\mathrm{b}}\,h^{2}}{0.02}\right)}^{3/2}\,
{\left(\frac{f_{\mathrm{g}}}{0.16}\right)}^{1/3}
\end{equation}
\noindent (Schaye 2001).
The parameter $h$ is the Hubble constant divided by 100, the \ion{H}{i}
photoionization rate $\Gamma \equiv \Gamma_{12} \times 10^{12}$
s$^{-1}$, the temperature of the Ly$\alpha$ forest $T \equiv T_{4}
\times 10^{4}$ K, 
$f(N_\ion{H}{i},z)$ is the differential density distribution function
as a function of $z$, $f_{\mathrm{g}}$ is a fraction of mass in the
Ly$\alpha$ forest and $\Omega_{\mathrm{b}}$ is the baryon density
(Schaye 2001). As the gas is assumed to be isothermal in Eq.~1,
$T$ is assumed to be constant ($\alpha = 0$ in 
$T \propto (1+\delta)^{\alpha}$), i.e. not a
function of $(1+\delta)$. The $d\Omega_{\mathrm{g}}/d\log(1+\delta)$
values, however, do not vary much with $\alpha$.

If $f(N_\ion{H}{i},z) \propto N_\ion{H}{i}^{-\beta}$,
$\Omega_{\mathrm{g}} \propto \int N_\ion{H}{i}^{4/3-\beta} \, d\ln
N_\ion{H}{i}$ (Schaye 2001). If the differential density distribution
function is a single power-law, then
$d\Omega_{\mathrm{g}}/d\log(1+\delta) \propto
N_\ion{H}{i}^{4/3-\beta}$.  Therefore, the shape of
$d\Omega_{\mathrm{g}}/d\log(1+\delta)$ reflects the deviation from the
single power-law of $f(N_\ion{H}{i},z)$.  As seen in
Fig.~\ref{fig_beta}, $\beta$ is not a constant.  If $\beta < 4/3$,
$d\Omega_{\mathrm{g}}/d\log(1+\delta)$ increases. Therefore,
$d\Omega_{\mathrm{g}}/d\log(1+\delta)$ is expected to increase at
smaller $N_\ion{H}{i}$ and to decrease at larger $N_\ion{H}{i}$.

\begin{figure}
\psfig{file=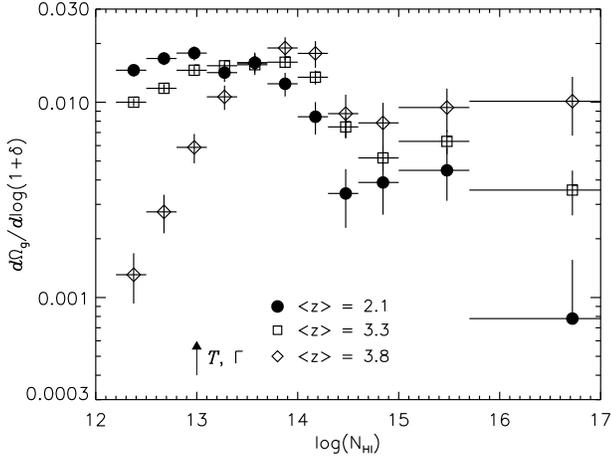,width=8.6cm}
\caption{The differential mass density distribution of the Ly$\alpha$
forest as a function of $N_\ion{H}{i}$. Among the parameters used in
Eq.~\ref{eq2} and Eq.~\ref{eq3}, $\Gamma_{12}$ was read from figure 8
of Haardt \& Madau (1996) and $T_{4}$ was read from
figure 3 of Schaye et al. (2000) for their simulations under
$J_{\mathrm{HM}}$ at $\delta = 1$.  The parameters
$f_{\mathrm{g}}$, $\Omega_{\mathrm{b}}\,h^{2}$ and $h$ are assumed to be
0.16, 0.02 and 0.65, respectively, in the flat Friedmann universe.  
At $<\!z\!> \  = 2.1$, 3.3 and 3.8,
$(T_{4},\Gamma_{12})$ is $(1.41,1.35)$, $(1.57,1.07)$
and $(1.62,0.7)$, respectively. The arrow indicates the direction
towards which $d\Omega_{\mathrm{g}}/d\log(1+\delta)$ moves if $T_{4}$
or $\Gamma_{12}$ increases.}
\label{mass}
\end{figure}

Fig.~\ref{mass} shows the differential mass density distribution
function $d\Omega_{\mathrm{g}}/d\log(1+\delta)$ as a function of
$N_\ion{H}{i}$ at three redshifts. The arrow in Fig.~\ref{mass}
indicates the direction towards which
$d\Omega_{\mathrm{g}}/d\log(1+\delta)$ moves if $T_{4}$ or
$\Gamma_{12}$ increase, i.e. preserves the shape of
$d\Omega_{\mathrm{g}}/d\log(1+\delta)$.  The QSO-dominated UV
background by Haardt \& Madau (1996), $J_{\mathrm{HM}}$, was assumed in
the Friedmann universe with $h = 0.65$.  
Other parameters are obtained from the observations, while $T_{4}$ 
was read from figure 3 by Schaye et al. (2000).

There is a turnover at $N_\ion{H}{i} \sim 10^{13} \ {\mathrm{cm}^{-2}}$
due to line blending as seen in Fig.~\ref{fig_ddf} and
in part due to the real deficiency of these weaker lines (cf. Schaye
2001). The amount of the
turnover is more significant at $z \sim 3.8$ and the turnover occurs at
lower $N_\ion{H}{i}$ at lower $z$. At $N_\ion{H}{i} \approx
10^{13.5-15} \ {\mathrm{cm}^{-2}}$,
$d\Omega_{\mathrm{g}}/d\log(1+\delta)$ is roughly approximated by a
single power-law, with a slope being slightly steeper at lower $z$.
This indicates that $\beta$ is larger at lower $z$ and that
$\beta$ at all $z$ is
larger than $4/3$.  At $N_\ion{H}{i} > 10^{15} \ {\mathrm{cm}^{-2}}$,
$d\Omega_{\mathrm{g}}/d\log(1+\delta)$ is not clearly defined due to a
larger size of the $N_\ion{H}{i}$ bin (also due to a smaller number of
lines at this column density regime).

The three curves for $d\Omega_{\mathrm{g}}/d\log(1+\delta)$ at
different redshifts agree with each other reasonably well with each
other if one applies a redshift-dependent offset in $N_\ion{H}{i}$, at
least in the range $N_\ion{H}{i} \approx 10^{13-15}
\ {\mathrm{cm}^{-2}}$ (referenced to $<z>=3.3$).  In fact,
$d\Omega_{\mathrm{g}}/d\log(1+\delta)$ as a function of $(1+\delta)$
does show a similar shape at different $z$.  Eq.~\ref{eq3} shows that
any change in $z$ and $\Gamma_{12}$ (thus $T_{4}$) reflects the
relation between $N_\ion{H}{i}$ and $(1+\delta)$. Thus, Fig.~\ref{mass}
indicates that the redshift-evolution of the differential density
distribution function seen at $N_\ion{H}{i} = 10^{13-15}
\ {\mathrm{cm}^{-2}}$ reflects the change in $\Gamma$ at different $z$
(Schaye 2001).

\begin{table*}
\caption[]{The power-law fit of the distribution functions,
$f(N_\ion{H}{i}) = A\,N_\ion{H}{i}^{-\beta}$}
\label{tab5}
\begin{tabular}{cccccccccccc}
\hline
\noalign{\smallskip}
& \multicolumn{2}{c}{$N_\ion{H}{i} = 10^{12.5-14} \ \mathrm{cm}^{-2}$}&
& \multicolumn{2}{c}{$N_\ion{H}{i} = 10^{14-16} \ \mathrm{cm}^{-2}$}&
& \multicolumn{2}{c}{$N_\ion{H}{i} = 10^{13-17} \ \mathrm{cm}^{-2}$}&
& \multicolumn{2}{c}{$N_\ion{H}{i} = 10^{14.5-17} \ \mathrm{cm}^{-2}$}\\
\\[-2ex]
\cline{2-3} \cline{5-6} \cline{8-9} \cline{11-12} \\[-2ex]
$<\!z\!>$ & $\log A$ & $\beta$ & &
$\log A$ & $\beta$  & & $\log A$ & $\beta$ & & $\log A$ & $\beta$\\
\noalign{\smallskip}
\hline
\noalign{\smallskip}
0.03$^{\mathrm{a}}$ & $11.3 \pm 0.7$ & $1.72 \pm 0.06$ &  
& $7.4 \pm 5.2$ & $1.43 \pm 0.35$ &  & - & - &  & - & - \\
0.17$^{\mathrm{b}}$ & - & - &  & - & - &  & $10.87 \pm 0.12$ 
& $2.04 \pm 0.23$ & & - & -\\
2.1 & $7.11 \pm 1.49$ & $1.42 \pm 0.03$ &
& $12.82 \pm 1.01$ & $1.87 \pm 0.11$ & 
& $11.00 \pm 1.38$ & $1.71 \pm 0.03$ &
& $10.11 \pm 0.79$ & $1.66 \pm 0.14$\\
3.3 & $3.26 \pm 1.33$ & $1.11 \pm 0.04$ &
& $12.21 \pm 1.04$ & $1.75 \pm 0.09$ &
& $9.18 \pm 1.31$ & $1.55 \pm 0.03$ &
& $11.40 \pm 0.82$ & $1.70 \pm 0.13$\\
3.8 & $1.94 \pm 1.17$ & $1.01 \pm 0.01$ & &
$10.56 \pm 0.98$ & $1.63 \pm 0.09$ &
& $7.77 \pm 1.22$ & $1.44 \pm 0.03$ &
& $7.15 \pm 0.81$ & $1.41 \pm 0.11$\\
\noalign{\smallskip}
\hline
\end{tabular}
\begin{list}{}{}
\item[$^{\mathrm{a}}$] Penton et al. (2000).
\item[$^{\mathrm{b}}$] Dav\'e \& Tripp (2001).
\end{list}
\end{table*}

\subsection{The two-point function of the flux}

The line width of the absorption lines (the $b$ parameter from the
Voigt profile fitting) provides valuable information on the thermal
temperature of the absorption clouds, if the Ly$\alpha$ forest is
thermally broadened. In particular, the lower envelope of the
$N_\ion{H}{i}$--$b$ diagram is interpreted to give an upper limit on
the temperature (through $b$) of the forest as a function of
$N_\ion{H}{i}$ (Schaye et al. 1999; 
McDonald et al. 2000; Ricotti, Gnedin \& Shull 2000;
Schaye et al. 2000; KCD; Kim, Cristiani \& D'Odorico 2002).  However,
it is not straightforward to determine this minimum $b$ value as a
function of $N_\ion{H}{i}$, particularly because of profile fitting
errors in blended features.

Instead of using the fitted parameters, the direct use of normalized
flux, $F$, has been proposed (Miralda-Escud\'e et al. 1997; Bryan et
al. 1999; Machacek et al. 2000; Theuns et al. 2000).  Among these
flux-based properties of the Ly$\alpha$ forest, the two-point function
of the flux provides the profile shape of absorption lines at $30 <
\Delta v < 100$ km s$^{-1}$ (Machacek et al. 2000; Theuns et al.
2000).

The two-point function of the flux, $P(F_{1},F_{2},\Delta v)$, is
the probability of two pixels with the $\Delta v$ velocity separation
having normalized fluxes $F_{1}$ and $F_{2}$. It is described as
\begin{eqnarray*}
\overline{\Delta F} (\Delta v,\delta F_{1})
\end{eqnarray*}
\begin{equation}
\equiv \int_{\delta F_{1}}\, \left[\int^{\infty}_{-\infty}
(F_{1}-F_{2})P(F_{1},F_{2},\Delta v)\,dF_{2}\right] \, dF_{1}/\delta F_{1},
\end{equation}
where $\overline{\Delta F} (\Delta v,\delta F_{1})$ is the mean flux
difference between two pixels with $F_{1}$ and $F_{2}$, which are
separated by $\Delta v$ (Theuns et al. 2000).

\begin{figure}
\psfig{file=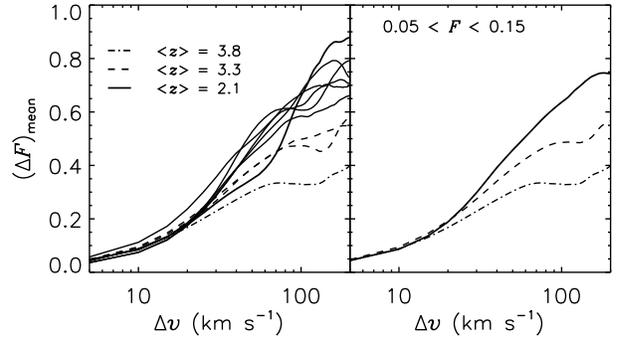, width=8.65cm}
\caption{The $\overline{\Delta F} (\Delta v,\delta F_{1})$ profile for
$0.05 < F < 0.15$ at different redshifts.  The left panel is for
individual QSOs of Sample A, while the right panel is averaged for each
$z$ bin.}
\label{fig_fdd1}
\end{figure}

Fig.~\ref{fig_fdd1} shows $\overline{\Delta F}$ for $0.05 < F < 0.15$
at different redshifts. We measured the $\Delta v (0.3)$ value,
following the definition by Machacek et al. (2000), the width of
$\overline{\Delta F} (\Delta v,\delta F_{1})$ at which
$\overline{\Delta F} (\Delta v,\delta F_{1})$ becomes 0.3. The
individual QSOs show a different $\Delta v (0.3)$ even at the same $z$.
The measured $\Delta v (0.3)$ averaged at each $z$ is 30.2 km s$^{-1}$,
33.7 km s$^{-1}$ and 48.8 km s$^{-1}$ at $<\!z\!> \ =$ 2.1, 3.3 and
3.8, respectively. This result might indicate 
that the line profile becomes
broader as $z$ increases.  In fact,
Theuns et al. (2000) note that at a given
$z$ a simulation with a hotter gas temperature shows a wider
$\overline{\Delta F} (\Delta v,\delta F_{1})$ profile than a simulation
with a lower gas temperature.

Decreasing $\Delta v (0.3)$ as decreasing $z$, however, is not an
indicative of decreasing in temperature.  
The reason of decreasing $\Delta v (0.3)$ is that 
$\overline{\Delta F} (\Delta v,\delta F_{1})$ must be asymptotic to 
$\overline{F} - 0.1$ and $\overline{F}$ decreases
as $z$ increases (Theuns et al. 2000). Since $\overline{F}$ depends
on $z$, comparing $\Delta v (0.3)$ at different $z$ directly
does not provide any information on the temperature. 

There is another way to look
at decreasing $\Delta v (0.3)$ with decreasing $z$. 
From the minimum cutoff $b$
distribution at each $z$ from the same data, Kim et al. (2002) show
that the minimum cutoff $b$ values increase as $z$ decreases (cf. Kim
et al. 1997). This result shows that the $b$ (i.e. line width) values
of absorption lines increases as $z$ decreases.  This result is
contrary to the result from $\overline{\Delta F} (\Delta v,\delta
F_{1})$.  This analysis of the minimum cutoff $b$ values suggests that
$\overline{\Delta F} (\Delta v,\delta F_{1})$ actually measures the
relative amounts of line blending and higher-$N_\ion{H}{i}$ forest,
when it is compared at {\it different} redshifts. The $\Delta v (0.3)$
values increase as $z$ increases due to more severe blending and more
higher-$N_\ion{H}{i}$ forest at higher $z$, not due to increasing
temperature of absorbing clouds. $\overline{\Delta F} (\Delta v,\delta
F_{1})$ for higher $F$ ranges, i.e. excluding higher column density
lines, does not show any $z$-dependence.

\subsection{The optical depth correlation function}

We analysed the clustering properties of the Ly$\alpha$ forest, using
the step optical depth correlation function (Cen et al. 1998; KCD). The
optical depth correlation function is less biased than the line
correlation functions since the fitted lines from Voigt profile fitting
are not unique.

The step optical depth correlation function $\xi_{\tau, {\rm s}}$ is
defined by

\begin{equation}
\xi_{\tau, {\rm s}} (\Delta v) \equiv \frac{<\tau_{\rm s}(v+\Delta v)
\tau_{\rm s}(v)>}{<\tau_{\rm s}>^{2}} -1,
\end{equation}

\noindent where the step optical depth, $\tau_{\mathrm{s}}$, is
$\tau_{\mathrm{s}} = 0$ if $\tau_{\mathrm{obs}} \le
\tau_{\mathrm{min}}$ and $\tau_{\mathrm{s}} = 1$ if
$\tau_{\mathrm{obs}} \ge \tau_{\mathrm{min}}$.

\begin{figure}
\psfig{file=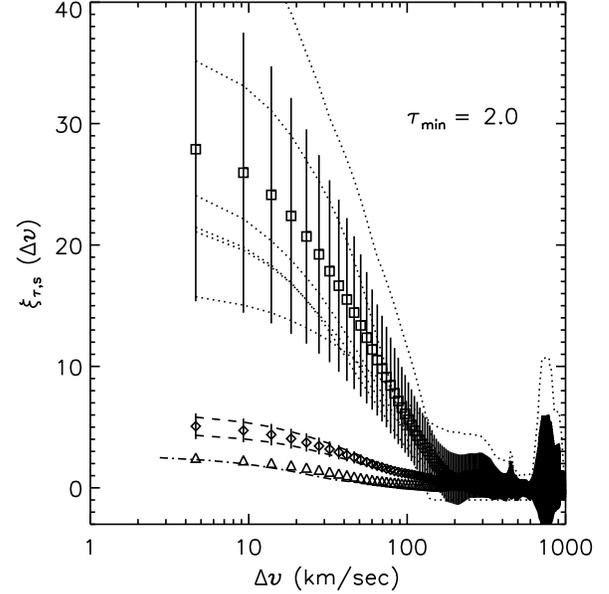,width=8.65cm}
\caption{The step optical depth correlation functions at $<\!z\!> \ =
2.1$
(squares), $<\!z\!> \ = 3.3$ (diamonds) and $<\!z\!> \ = 3.8$
(triangles).
Error bars are the $1\sigma$
statistical errors from the sample size.}
\label{fig_step}
\end{figure}

Fig.~\ref{fig_step} shows $\xi_{\tau, {\rm s}} (\Delta v)$ averaged at
each $z$ for $\tau_{\mathrm{min}} = 2$. The $\tau_{\mathrm{min}} = 2$
corresponds to $N_\ion{H}{i} \sim 10^{13.6} \ {\mathrm{cm}}^{-2}$, if
$b$ is assumed to be $b=30$ km s$^{-1}$. Dotted lines, dashed lines and
the dot-dashed line represent the individual QSOs in each $z$ bin at
$<\!z\!> \ =$ 2.1, 3.3 and 3.8.  The large error bars at $<\!z\!> \ =
2.1$ are due to the fact that the correlation strengths of different
QSOs at similar redshifts show a wide range. Although there is a large
scatter at $<\!z\!> \ =$ 2.1, the step optical depth correlation
functions show a strong clustering at $v < 100$ km s$^{-1}$.  In
addition, the step optical depth correlation strength increases as $z$
decreases in general.  At $<\!z\!> \ =2.1$, $\xi_{\tau, {\rm s}}$ (50
km s$^{-1}$) shows a $\sim 10\sigma$ significance compared
to that at $<\!z\!> \ =3.3$.  This stronger clustering of the
Ly$\alpha$ forest at $z \le 2$ compared to higher $z$ is expected from
the differences in $dn/dz$ and $\overline{\tau}_\ion{H}{i}$ along the
different sightlines.  Studies of the two-point velocity correlation
strength using the fitted line parameters at $z < 1.5$ and $z > 2.5$
also show a velocity correlation at $\Delta v \approx$ 50--500 km
s$^{-1}$, although its significance is much smaller than that of the
step optical depth correlation function (Ulmer 1996; Cristiani et al.
1997; Penton et al.  2000).

\section{The Ly$\beta$ forest}

At $N_\ion{H}{i} \ge 10^{14} \ {\mathrm{cm}^{-2}}$, the Ly$\alpha$
forest absorption lines start to saturate. For single lines, this
saturation results in an increasingly inaccurate estimation of
$N_\ion{H}{i}$ and $b$ as the $\ion{H}{i}$ column density increases, at
least up to the stage where damping wings become significant.  More
importantly, saturation also results in the uncertainty in the number
of subcomponents present.  In reality, some high-$N_\ion{H}{i}$ clouds
are not a really high-$N_\ion{H}{i}$ cloud, but a blend of several
lower-$N_\ion{H}{i}$ clouds. Many of these saturated lines could have
more accurate measurements of $N_\ion{H}{i}$, $b$ and a number of
subcomponents if the higher order Lyman lines are fitted simultaneously
with the Ly$\alpha$, since the oscillator strengths decrease
monotonically as one goes up the Lyman series.

Fitting the absorption profiles with higher-order series becomes more
time-consuming as $z$ increases. In some cases, severe contamination by
lower-$z$ Ly$\alpha$ forest and by the other higher-order Lyman series
from higher-$z$ forest make it very difficult to deblend the saturated
Ly$\alpha$ lines properly.  In addition, S/N of $\ge 15$--20 is
required to determine the continuum with adequate reliability.  For
these reasons, we have only fitted two QSO spectra with S/N $\sim$
10--30 in the Ly$\beta$ regions, HE1122--1648 and
HE2217--2818\footnote{Note that these two spectra have S/N of 20--30 in
3200--3500 \AA\/. Shorter than 3200 \AA\/, S/N decreases very rapidly,
S/N $\sim$ 10--20. We are, however, mainly interested in constraining a
reliable $N_\ion{H}{i}$ determination of saturated $N_\ion{H}{i}$
clouds in the Ly$\alpha$ forest regions and the higher-$N_\ion{H}{i}$
clouds in the Ly$\beta$ forest regions. Therefore, lower S/N at
3100--3200 \AA\/ does not limit to study the Ly$\beta$ forest and the
higher-$N_\ion{H}{i}$ Ly$\alpha$ forest in the Ly$\beta$ regions.}.

\subsection{The line number density evolution of the Ly$\beta$ forest}

\begin{figure}
\begin{center}
\psfig{file=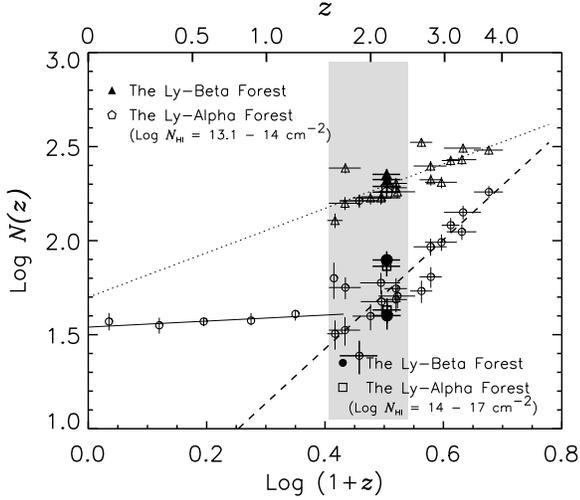, width=8.65cm}
\end{center}
\caption{The line number density evolution of the Ly$\alpha$ forest
and the Ly$\beta$ forest. Open circles represent all the different
symbols from Fig.~\ref{fig_dndz2}, while open triangles are from
Fig.~\ref{fig_dndz4}. The dashed line and the dotted line represent the
maximum-likelihood fit of the Ly$\alpha$ forest
for each $N_\ion{H}{i}$ range in the fit.}
\label{fig_dndz5}
\end{figure}

Fig.~\ref{fig_dndz5} shows the line number density, $dn/dz$, of the
Ly$\beta$ forest (filled symbols) and the Ly$\alpha$ forest both from
Sample A and Sample B (open symbols).  For $N_\ion{H}{i} = 10^{14-17}
\ {\mathrm{cm}}^{-2}$, $dn/dz$ of the Ly$\beta$ forest is within the
error bars of $dn/dz$ of the Ly$\alpha$ forest. In the case of
HE2217--2818, $dn/dz$ of the Ly$\beta$ forest is less than $1\sigma$
different from that of the Ly$\alpha$ forest (14 versus 15 lines). For
HE1122--1648, it becomes 27 versus 25 lines, well within the error
bars.  On the other hand, for $N_\ion{H}{i} = 10^{13.1-14}
\ {\mathrm{cm}}^{-2}$, $dn/dz$ of the Ly$\beta$ forest systems shows a
larger deviation.  For HE2217--2818 (for HE1122--1648), it is 74 versus
68 lines (77 versus 62 lines). The main reasons for this difference in
$dn/dz$ come in part from the continuum re-adjustment in the
Ly$\alpha$-Ly$\beta$ fit (which has a more significant effect on the
lower $N_\ion{H}{i}$ forest) and in part from the more correct
deblending of the higher-$N_\ion{H}{i}$ forest. When the
higher-$N_\ion{H}{i}$ forest is deblended into more components, the
number of lower-$N_\ion{H}{i}$ components in the forest increases.
This results in the higher number of smaller $N_\ion{H}{i}$ forest from
using the line parameters determined from fitting the Ly$\alpha$-Ly$\beta$ forest.

In general, however, the uncertainty in $dn/dz$ of the Ly$\beta$ forest
is in the same amount or less than the uncertainty resulted from the
different sightlines, i.e. the cosmic variance. The determination of
$dn/dz$ from the Ly$\alpha$ forest {\it alone} does not affect the
overall results of $dn/dz$ at least at $1.5 < z < 2.5$. At higher $z$,
more severe line blending is likely to produce higher uncertainties in
the reliable number of subcomponents in the saturated lines. This might
result in a significantly different $dn/dz$ of the Ly$\beta$ forest at
$z > 3$, i.e. a higher $dn/dz$, compared to that of the Ly$\alpha$
forest.

\subsection{The differential density distribution function of the
Ly$\beta$ forest}

\begin{figure}
\psfig{file=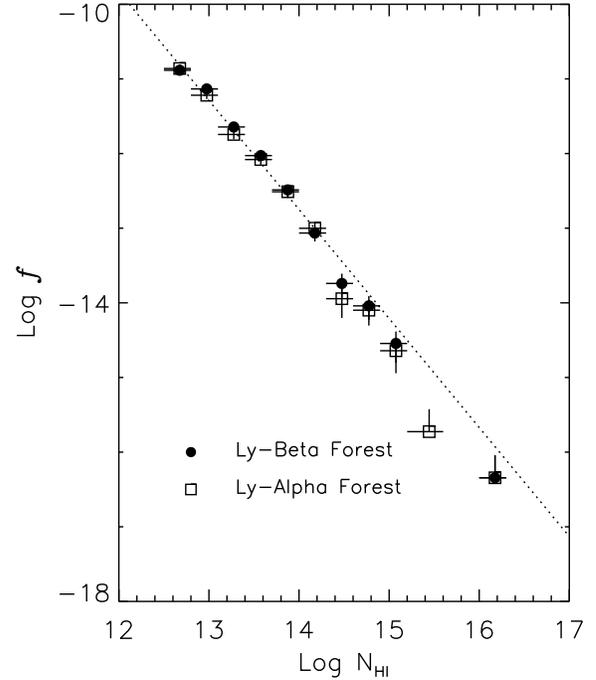,width=8.65cm}
\caption{The differential column density distribution function from
the Ly$\alpha$-only fit and the Ly$\alpha$ and Ly$\beta$ fit. The
data are shown in the binned sample for the display purpose.
The dotted line represents the incompleteness-corrected
$f(N_\ion{H}{i})$ at $<\!z\!> \ = 2.85$ from Hu et al. (1995),
$f(N_\ion{H}{i}) = 4.9 \times 10^{7} \, N_\ion{H}{i}^{-1.46}$.}
\label{fig_ddf2}
\end{figure}

Fig.~\ref{fig_ddf2} shows the differential column density distribution
function, $f(N_\ion{H}{i})$, for the Ly$\alpha$-only fit (open squares)
and for the Ly$\alpha$-Ly$\beta$ fit (filled circles).  At
$N_\ion{H}{i} \le 10^{14.3} \ {\mathrm{cm}}^{-2}$, there is no
significant difference in the Ly$\beta$ forest and the Ly$\alpha$
forest.  The Kolmogorov-Smirnov test gives the KS statistic of 0.06 and
the probability of 0.56.  Note that a slight difference in
$f(N_\ion{H}{i})$ is due to the continuum fitting adjustment when the
Ly$\alpha$ and the Ly$\beta$ forest lines were fitted simultaneously.

The difference seems to become more noticeable at $N_\ion{H}{i} \ge
10^{14.3} \ \mathrm{cm}^{-2}$. This is, however, in part caused by the
fact that there are not many high-$N_\ion{H}{i}$ clouds in the sample (22
lines in the Ly$\beta$ forest sample and 18 lines in the Ly$\alpha$
forest sample) and in part by the fact that Fig.~\ref{fig_ddf2} shows
the binned data for the display purpose. In fact, the
Kolmogorov-Smirnov test gives the KS statistic of 0.22 and the
probability of 0.65.  Although it is a statistically small sample,
Table~\ref{tab6} lists $f(N_\ion{H}{i})$ at different $N_\ion{H}{i}$
ranges. Within the error bars, $f(N_\ion{H}{i})$ from both the
Ly$\alpha$ forest and the Ly$\beta$ forest is in good agreement.

\begin{table}
\caption[]{The power-law fit of the distribution functions,
$f(N_\ion{H}{i}) = A\,N_\ion{H}{i}^{-\beta}$ at $<\!z\!> \ = 2.2$}
\label{tab6}
\begin{tabular}{cccccc}
\hline
\noalign{\smallskip}
& \multicolumn{2}{c}{$N_\ion{H}{i} = 10^{12.5-14.3} \ \mathrm{cm}^{-2}$}&
&
\multicolumn{2}{c}{$N_\ion{H}{i} = 10^{14.3-17} \ \mathrm{cm}^{-2}$} \\
\\[-2ex]
\cline{2-3} \cline{5-6} \\[-2ex]
& $\log A$ & $\beta$ & &
$\log A$ & $\beta$  \\
\noalign{\smallskip}
\hline
\noalign{\smallskip}
Ly$\alpha$ & $7.40 \pm 1.31$ & $1.43 \pm 0.04$ &
& $11.57 \pm 0.69$ & $1.74 \pm 0.17$ \\
Ly$\beta$ & $7.52 \pm 1.34$ & $1.43 \pm 0.04$ &
& $11.92 \pm 0.75$ & $1.76 \pm 0.15$ \\
\noalign{\smallskip}
\hline
\end{tabular}
\end{table}

\subsection{The distribution of the Doppler parameters of the
Ly$\beta$ forest}

One of the important physical properties of the Ly$\alpha$ forest, the
temperature of the absorbing gas, can be derived from the distribution
of the Doppler ($b$) parameters.  At higher $z$, the absorption lines
are broadened by the thermal motion as well as the bulk motions.
Therefore, the lower cutoff envelope of the $N_\ion{H}{i}$--$b$
distribution constrains an upper limit on the thermal temperature of
the forest as a function of $N_\ion{H}{i}$ at a given $z$ (Hui \&
Gnedin 1997; Schaye et al. 1999; McDonald et al. 2000; Ricotti et al.
2000; KCD; Dav\'e \& Tripp 2001; Kim et al. 2002).

\begin{figure}
\psfig{file=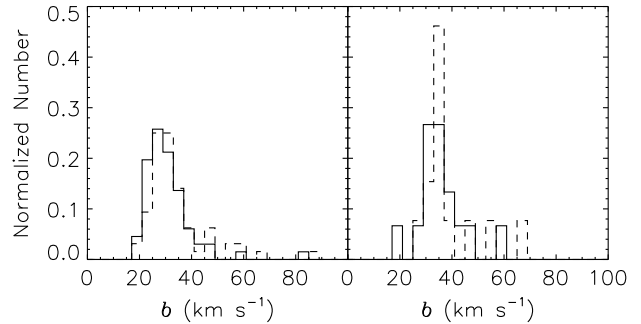,width=8.65cm}
\vspace{-2.5cm}
\caption{The distribution of the Doppler parameters for the Ly$\alpha$
and Ly$\beta$ forest. The left hand panel is for
$N_\ion{H}{i}=10^{12.5-14.5} \ {\mathrm{cm}}^{-2}$, while the
right hand panel is for $N_\ion{H}{i}=10^{14.5-17} \
{\mathrm{cm}}^{-2}$. Solid lines are for the Ly$\beta$ forest and
dashed lines are for the Ly$\alpha$ forest.}
\label{fig_dop}
\end{figure}

Fig.~\ref{fig_dop} shows the distribution of the Doppler parameters of
the Ly$\alpha$ (dashed lines) and the Ly$\beta$ forest (solid lines).
For $N_\ion{H}{i}=10^{12.5-14.5} \ {\mathrm{cm}}^{-2}$ (the left hand
panel), the median $b$ is 27.6 km s$^{-1}$ (the Ly$\beta$ forest) and
26.84 km s$^{-1}$ (the Ly$\alpha$ forest). There is no significant
difference in the $b$ distribution between the Ly$\alpha$ forest and
the Ly$\beta$ forest for this $N_\ion{H}{i}$ range. This result
confirms the validity of the determination of the lower cutoff $b$
values using the Ly$\alpha$ forest in previous studies since the lower
cutoff has been derived at $N_\ion{H}{i} = 10^{12.5-14.5}
\ {\mathrm{cm}}^{-2}$.  Since these lines are not saturated, lower
cutoff $b$ values from the Ly$\alpha$ forest would not be changed, when
the Ly$\alpha$ forest is fit simultaneously with the Ly$\beta$ forest.
In fact, the $N_\ion{H}{i}$--$b$ diagram from the Ly$\alpha$ forest and
the Ly$\beta$ forest does not show any significant difference.

The right hand panel of Fig.~\ref{fig_dop} shows the $b$ distributions
for $N_\ion{H}{i}=10^{14.5-17} \ {\mathrm{cm}}^{-2}$.  The median $b$
is 34.0 km s$^{-1}$ (the Ly$\beta$ forest; 15 lines) and 32.7 km
s$^{-1}$ (the Ly$\alpha$ forest; 13 lines).  Although the median values
are not statistically robust due to the small number of available
lines, there is no significant difference between the Ly$\alpha$ and
the Ly$\beta$ forest for this higher $N_\ion{H}{i}$ range, either.

Table~\ref{tab7} lists the observed $\overline{b}$ values and the
median $b$ values from the Ly$\beta$ forest as well as the ones at $z
\sim 0.15$. Our characteristic $b$-values at $z \sim 2.2$ are very
similar to those reported by Shull et al. (2000) from the
curve-of-growth analysis. On the other hand, our values are somewhat
larger than those of Dav\'e \& Tripp (2001) obtained from a profile
fitting analysis. Table~\ref{tab7} shows that there is no significant
difference of $\overline{b}$ and median $b$ values between the
Ly$\alpha$ forest and the Ly$\beta$ forest at $z \sim 2.2$. Although
$\overline{b}$ and median $b$ values are dependent on $N_\ion{H}{i}$
ranges, Table~\ref{tab8} suggests that the median $b$ value does not
increase, i.e. constant or decreasing, at $z < 1.5$ from at $z \sim
2.1$ for two $N_\ion{H}{i}$ ranges considered in Table~\ref{tab8} (see
KCD; Kim et al. 2002).

\begin{table}
\caption{The characteristic $b$ values of the Ly$\beta$ forest}
\label{tab7}
\begin{tabular}{ccccc}
\hline
\noalign{\smallskip}
$<\!z\!>$ & $N_\ion{H}{i}$ & $b_{\mathrm{median}}$ & ${\overline{b}}$ & Ref. \\
& $({\mathrm{cm}}^{-2})$ & (km s$^{-1}$) & (km s$^{-1}$) & \\
\noalign{\smallskip}
\hline
\noalign{\smallskip}
0.15 & $10^{13.76-17, \, {\mathrm{a}}}$ & $28$ & $31.4 \pm 7.4$ & 1\\ 
2.2  & $10^{13.76-17}$ & $28.5$ & $30.2$ & 2\\ 
0.17 & $10^{13-17}$ & $22$ & $25$ & 3 \\
2.2 & $10^{13-17}$ & $27.0$ & $32.6$ & 2 \\
\noalign{\smallskip}
\hline
\end{tabular}
\begin{list}{}{}
\item[$^{\mathrm{a}}$] The equivalent width threshold $W \ge 0.2$\AA\/
has been translated to $N_\ion{H}{i} = 10^{13.76} \ {\mathrm{cm}}^{-2}$,
assuming $b=30$ km s$^{-1}$.
\item[Ref.:] 1. Shull et al. (2000); 2. This study;
3. Dav\'e \& Tripp (2001).
\end{list}
\end{table}

\begin{table}
\caption{The characteristic $b$ values of the Ly$\alpha$ forest}
\label{tab8}
\begin{tabular}{ccccc}
\hline
\noalign{\smallskip}
$<\!z\!>$ & $N_\ion{H}{i}$ & $b_{\mathrm{median}}$ & ${\overline{b}}$ &
Ref. \\
& $({\mathrm{cm}}^{-2})$ & km s$^{-1}$ & km s$^{-1}$ & \\
\noalign{\smallskip}
\hline
\noalign{\smallskip}
0.15 & $10^{13.76-17, \,{\mathrm{a}}}$ & $28$ & $31.4 \pm 7.4$ & 1\\
2.1  & $10^{13.76-17}$ & $29.5$ & $32.7$ & 2\\
3.3  & $10^{13.76-17}$ & $27.5$ & $30.7$ & 2\\
3.8  & $10^{13.76-17}$ & $31.0$ & $39.5$ & 3\\
0.17 & $10^{13-17}$ & $22$ & $25$ & 4 \\
2.1 & $10^{13-17}$ & $27.0$ & $31.5$ & 2\\
3.3 & $10^{13-17}$ & $28.1$ & $31.6$ & 2\\
3.8 & $10^{13-17}$ & $29.2$ & $35.9$ & 3\\
\noalign{\smallskip}
\hline
\end{tabular}
\begin{list}{}{}
\item[$^{\mathrm{a}}$] The equivalent width threshold $W \ge 0.2$\AA\/
has been translated to $N_\ion{H}{i} = 10^{13.76} \ {\mathrm{cm}}^{-2}$,
assuming $b=30$ km s$^{-1}$.
\item[Ref.:] 1. Shull et al. (2000);
2. This study;
3. Lu et al. (1996);
4. Dav\'e \& Tripp (2001).
\end{list}
\end{table}

\section{Conclusions}

We have analyzed the properties of the Ly$\alpha$ forest at
$N_\ion{H}{i} = 10^{12.5 -17} \mathrm{cm}^{-2}$ toward 8 QSOs, using
high resolution ($R \sim 45\,000$), high S/N ($\sim$ 25--40) VLT/UVES
data. The analyses presented here extend the
ones by KCD, using the data obtained and treated uniformly. In
addition, the 6 QSOs at $1.5 < z < 2.5$ enable us to explore the
cosmic variances along different sightlines. 
Combined with other high-resolution observations from the literature
as well as {\it HST} observations at $z < 1.5$, we have studied the
properties of the Ly$\alpha$ forest as a function of $z$.

To analyse the properties of the Ly$\alpha$ forest, we have used
two profile fitting approaches. In the first analysis, we 
have only fitted
the Ly$\alpha$ absorption profiles in between the Ly$\alpha$ and the
Ly$\beta$ emission lines. In the second analysis, 
we have fitted the
Ly$\alpha$ forest simultaneously with the higher order lines of the Lyman
series down to 3050 \AA\/, for two QSOs at $z_{\mathrm{em}} \sim 2.4$.
The second analysis has been adopted to
probe the properties of the Ly$\beta$ forest at $z \sim 2.2$ and
to investigate the general differences between studies from the 
traditional Ly$\alpha$-only fits and from the higher orders of the
Lyman series fits.
In addition, we have also applied the optical depth analysis.
For the lines with $N_{\ion{H}{i}} = 10^{12.5-17} \ 
{\mathrm{cm}}^{-2}$, we have in general confirmed the conclusions by KCD
derived from a smaller sample than that of this study.
We have found:

1) The line number density of the Ly$\alpha$ forest, $dn/dz$, is fit
well by a single power-law and shows a steeper evolution at 
higher-$N_\ion{H}{i}$ forest at $1.5 < z < 4$.  For $N_\ion{H}{i} =
10^{13.64-17} \ \mathrm{cm}^{-2}$, $dn/dz \propto (1+z)^{2.47 \pm
0.18}$. For $N_\ion{H}{i} = 10^{14-17} \ \mathrm{cm}^{-2}$, $dn/dz
\propto (1+z)^{2.90 \pm 0.25}$.  The change in the 
evolutionary behavior of $dn/dz$ occurs at $z
\sim 1$.  
In addition, forest regions with
under-abundant line density and regions with over-abundant line density
start to appear at $z < 2.5$.  This small-scale variation is more
significant for higher $N_\ion{H}{i}$ forest. This is probably caused
by the small-scale structure evolution in the forest, i.e. the
increasing clustering of high $N_\ion{H}{i}$ forest at $z < 2.5$.

2) The mean \ion{H}{i} opacity, ${\overline{\tau}}$, is also well
approximated by a single power law at $1.5 < z < 4$,
${\overline{\tau}}_\ion{H}{i} \propto (1+z)^{3.37 \pm 0.20}$. This
result is about a factor of 1.3 smaller than the commonly used
${\overline{\tau}}$ determined by Press et al. (1993) at all $z$.  When
compared with ${\overline{\tau}}$ at $z < 1.5$ from {\it HST}
observations, ${\overline{\tau}}$ at $z \sim 0$ is at least a factor of
4 higher than the one extrapolated from at $z > 1.5$.  The different
evolution in $\overline{\tau}_\ion{H}{i}$ occurs at $z < 1$, a bit
lower than that suggested from $dn/dz$.

3) For $N_\ion{H}{i} = 10^{12.5-15} \ {\mathrm{cm}^{-2}}$,
the differential column density
distribution function, $f(N_\ion{H}{i})$,
can be best fit by $f(N_\ion{H}{i}) \propto N_{\ion{H}{i}}^{-\beta}$
with $\beta \approx 1.5$. The $\beta$ values, however, shows the dependence
on the $N_\ion{H}{i}$ range in the fit, mostly due to line
blending at $N_\ion{H}{i} \le 10^{13.5} \ {\mathrm{cm}}^{-2}$.
When combined with {\it HST} observations,
the exponent $\beta$ increases as $z$ decreases at $0 < z < 4$
for $N_\ion{H}{i} = 10^{13-17} \
{\mathrm{cm}^{-2}}$.

4) The two-point function of the flux does not represent a shape of
line profiles as previously suggested. Instead, when it is compared at
different redshifts, it represents a degree of line blending in the
forest.

5) The step optical correlation function confirms that lines with
higher opacity (strong lines) are more clustered than lines with lower
opacity (weak lines) at the velocity of $\le 200$ km s$^{-1}$.

6) The analyses of the Ly$\beta$ forest at $z \sim 2.2$ are in good
agreement with those of the Ly$\alpha$ forest. This result shows that
previous studies on the Ly$\alpha$ forest-only have not been
significantly biased from line blending and saturation at least at $z <
2.5$. Line blending, however, could be more problematic at higher $z$
when it becomes more severe, changing some of the results from the
analysis of higher-$N_\ion{H}{i}$ Ly$\alpha$ forest at $N_\ion{H}{i}
\ge 10^{14.5} \ {\mathrm{cm}}^{-2}$.

\section{Acknowledgments}
We are indebted to all people involved in the conception, construction,
commissioning and science verification of UVES and UT2 for the quality
of the data used in this paper.  We are grateful to Simone Bianchi
and Joop Schaye for
their helpful discussions. RFC is grateful to ESO for support through their 
visitor programme. This work has been conducted with partial
support by the Research Training  Network "The Physics of the
Intergalactic Medium" set up by the European Community under the
contract HPRN-CT2000-00126 RG29185 and by ASI through contract
ARS-98-226.

\appendix

\section{The line list of Q1101--264}

This is an example of a part of the line lists, which are published
electronically at the CDS site to cdsarc.u-strasbg.fr 
(130.79.128.5).

\begin{table*}
\caption{The line list of Q1101--264}
\label{tab11}
\begin{tabular}{rclcrrcc}
\hline
\noalign{\smallskip}
 \# & $\lambda$ & Identification & $z$ & $b \;\;$ & $\sigma(b)$ &
$\log N_\ion{H}{i}$ & $\sigma(\log N_\ion{H}{i})$ \\
& (\AA\/) & & & km s$^{-1}$ & km s$^{-1}$ & cm$^{-2}$ & cm$^{-2}$ \\
\noalign{\smallskip}
\hline
\noalign{\smallskip}
   0 & 3233.622  & H I          & 1.65995  &  38.76 & 13.56 & 12.291 & 0.124 \\
   1 & 3234.807  & H I          & 1.66092  &  43.36 &  9.53 & 12.554 & 0.073 \\
   2 & 3237.851  & Fe II 2382.8 & 0.35886  &  17.06 &  7.92 & 12.002 & 0.207 \\
   3 & 3238.092  & Fe II 2382.8 & 0.35896  &   5.95 &  0.23 & 13.443 & 0.019 \\
   4 & 3238.304  & H I          & 1.66380  &   9.46 &  3.81 & 12.115 & 0.109 \\
   5 & 3238.311  & Fe II 2382.8 & 0.35906  &   7.14 &  4.99 & 11.806 & 0.186 \\
   6 & 3238.499  & Fe II 2382.8 & 0.35913  &   4.89 &  0.50 & 12.716 & 0.021 \\
   7 & 3238.673  & Fe II 2382.8 & 0.35921  &   5.95 &  2.25 & 11.979 & 0.087 \\
   8 & 3238.713  & H I          & 1.66414  &  15.04 &  5.94 & 12.100 & 0.124 \\
   9 & 3240.386  & H I          & 1.66551  &  30.22 &  3.03 & 12.558 & 0.036 \\
  10 & 3245.307  & H I          & 1.66956  &  26.69 &  1.47 & 12.734 & 0.020 \\
  11 & 3249.002  & Fe II 1144.9 & 1.83770  &   5.63 &  0.85 & 11.480 & 0.036 \\
  12 & 3249.567  & Fe II 1144.9 & 1.83820  &   7.99 &  2.73 & 11.826 & 0.153 \\
  13 & 3249.705  & H I          & 1.67318  &  25.80 & 13.77 & 12.283 & 0.265 \\
  14 & 3249.706  & Fe II 1144.9 & 1.83832  &   5.01 &  2.79 & 11.660 & 0.217 \\
  15 & 3249.961  & Fe II 1144.9 & 1.83854  &   5.86 &  0.19 & 12.836 & 0.011 \\
  16 & 3250.156  & Fe II 1144.9 & 1.83871  &   8.48 &  0.62 & 12.663 & 0.020 \\
  17 & 3250.212  & H I          & 1.67360  &  23.72 &  1.35 & 13.258 & 0.028 \\
  18 & 3250.369  & Fe II 1144.9 & 1.83890  &   5.41 &  0.12 & 13.161 & 0.008 \\
  19 & 3250.641  & Fe II 1144.9 & 1.83914  &   4.84 &  0.33 & 12.500 & 0.016 \\
  20 & 3250.758  & Fe II 1144.9 & 1.83924  &   1.04 &  0.99 & 12.598 & 0.564 \\
  21 & 3250.861  & Fe II 1144.9 & 1.83933  &   6.01 &  1.61 & 12.056 & 0.079 \\
  22 & 3251.115  & H I          & 1.67434  &  13.61 &  3.34 & 12.217 & 0.080 \\
  23 & 3252.117  & H I          & 1.67516  &  22.86 &  1.07 & 12.932 & 0.017 \\
  24 & 3252.864  & H I          & 1.67578  &  22.64 &  1.06 & 12.916 & 0.017 \\
  25 & 3256.105  & H I          & 1.67844  &  48.81 & 22.46 & 12.651 & 0.209 \\
  26 & 3256.790  & H I          & 1.67901  &  24.49 &  0.64 & 13.910 & 0.013 \\
  27 & 3257.472  & H I          & 1.67957  &  26.80 & 15.91 & 12.204 & 0.212 \\
  28 & 3261.481  & H I          & 1.68287  &  14.42 &  2.96 & 12.935 & 0.161 \\
  29 & 3261.695  & H I          & 1.68304  &  33.55 &  0.60 & 13.999 & 0.012 \\
  30 & 3262.569  & H I          & 1.68376  &  33.94 &  1.72 & 13.458 & 0.034 \\
  31 & 3262.807  & H I          & 1.68396  &  15.99 &  2.29 & 12.909 & 0.129 \\
  32 & 3265.423  & H I          & 1.68611  &  51.19 & 20.10 & 12.551 & 0.135 \\
  33 & 3266.246  & H I          & 1.68679  &  19.93 &  1.99 & 13.370 & 0.138 \\
  34 & 3266.566  & H I          & 1.68705  &  25.48 & 12.42 & 12.869 & 0.417 \\
  35 & 3271.193  & H I          & 1.69086  &  32.81 &  2.32 & 12.693 & 0.025 \\
  36 & 3273.691  & H I          & 1.69291  &  33.03 &  2.71 & 12.765 & 0.030 \\
  37 & 3275.972  & H I          & 1.69479  &  28.05 &  0.55 & 13.536 & 0.008 \\
  38 & 3278.837  & H I          & 1.69714  &  68.02 & 21.70 & 12.852 & 0.165 \\
  39 & 3279.695  & H I          & 1.69785  &  30.42 &  3.45 & 13.047 & 0.087 \\
  40 & 3280.860  & H I          & 1.69881  & 119.52 & 63.51 & 12.912 & 0.201 \\
  41 & 3283.691  & H I          & 1.70114  &  52.86 & 12.22 & 12.709 & 0.083 \\
  42 & 3284.505  & H I          & 1.70181  &  23.82 &  1.58 & 12.985 & 0.036 \\
  43 & 3288.110  & H I          & 1.70477  &  59.28 & 51.22 & 12.563 & 0.629 \\
  44 & 3288.429  & H I          & 1.70503  &  14.12 &  3.50 & 12.702 & 0.187 \\
  45 & 3288.903  & H I          & 1.70542  &  40.31 &  1.74 & 13.971 & 0.028 \\
  46 & 3289.781  & H I          & 1.70615  &  20.11 &  0.67 & 13.497 & 0.013 \\
  47 & 3290.299  & H I          & 1.70657  &  16.60 &  2.30 & 12.623 & 0.049 \\
  48 & 3292.783  & H I          & 1.70862  &  31.87 &  8.69 & 12.314 & 0.090 \\
  49 & 3295.336  & H I          & 1.71072  &  31.35 & 18.35 & 12.089 & 0.189 \\
  50 & 3296.100  & H I          & 1.71134  &  20.07 &  1.39 & 12.857 & 0.025 \\
  51 & 3296.846  & H I          & 1.71196  &  10.31 &  3.02 & 12.078 & 0.092 \\
  52 & 3298.412  & H I          & 1.71325  &  48.41 & 15.02 & 12.435 & 0.100 \\
  53 & 3299.860  & H I          & 1.71444  &  23.02 &  0.70 & 13.511 & 0.012 \\
  54 & 3301.287  & H I          & 1.71561  &  17.77 & 14.04 & 12.464 & 0.789 \\
  55 & 3301.603  & H I          & 1.71587  &  18.48 & 10.93 & 12.975 & 0.450 \\
  56 & 3302.003  & H I          & 1.71620  &  30.32 &  5.02 & 13.362 & 0.105 \\
  57 & 3304.878  & H I          & 1.71856  &  26.21 &  3.68 & 12.421 & 0.050 \\
  58 & 3306.464  & H I          & 1.71987  &  23.22 &  2.82 & 12.618 & 0.043 \\
  59 & 3309.185  & H I          & 1.72211  &  35.11 &  8.17 & 12.313 & 0.079 \\
\noalign{\smallskip}
\hline
\end{tabular}
\end{table*}

\end{document}